\documentclass{article}

\usepackage{arxiv}

\usepackage{hyperref}

\usepackage{graphicx}
\usepackage{mathptmx}

\usepackage{subfig}
\usepackage{amsmath}
\usepackage{amsfonts}
\usepackage{amssymb}
\usepackage{amsbsy}
\usepackage{amsthm}
\usepackage{xspace}
\usepackage{cases}
\usepackage{multicol}
\usepackage{tikz}
\usepackage[export]{adjustbox}

\usepackage{xfrac}
\usepackage{sidecap}
\usepackage{systeme}
\usepackage{etoolbox}

\usepackage{listings}
\usepackage{caption}
\usepackage{color}

\usepackage[boxruled,linesnumbered,titlenumbered]{algorithm2e}

\RestyleAlgo{ruled}

\usepackage{floatrow}   

\DeclareCaptionFormat{listing}{\rule{\dimexpr\textwidth-0.52cm\relax}{0.4pt}\par\vskip1pt#1:#2#3}
\newcommand*{\mathcolor}{}
\def\mathcolor#1#{\mathcoloraux{#1}}
\newcommand*{\mathcoloraux}[3]{%
  \protect\leavevmode
  \begingroup
    \color#1{#2}#3%
  \endgroup
}

\newcommand {\nxt}{\eta}

\newcommand{\abs}[1]{|#1|}
\newcommand{\blue}[1]{#1}

\newcommand\BibTeX{{\rmfamily B\kern-.05em \textsc{i\kern-.025em b}\kern-.08em
T\kern-.1667em\lower.7ex\hbox{E}\kern-.125emX}}

\newcommand{\dQRel}{\ensuremath{\Delta}\textrm{QRel}}
\newcommand{\dQMin}{\ensuremath{\Delta}\textrm{QMin}}
\newcommand{\deltaQ}{\ensuremath{\Delta}\textrm{Q}}

\newcommand{\uModelica}{$\mu$-Modelica}
\newcommand{\retQSS}{retQSS\xspace}

\newcommand{\PartPos}[1]{\ensuremath{\mathbf{x}_{#1}}}
\newcommand{\PartVel}[1]{\ensuremath{\mathbf{v}_{#1}}}
\newcommand{\PartDir}[1]{\ensuremath{\mathbf{d}_{#1}}}
\newcommand{\Flock}[1]{\ensuremath{\textrm{Flock}(#1)}}
\newcommand{\vecFromTo}[2]{\ensuremath{\PartPos{#2} - \PartPos{#1}}}

\newcommand{\ZC}{\ensuremath{\textrm{ZC}}}
\newcommand{\Handler}{\ensuremath{\textrm{H}}}

\newcommand{\MDErr}{\ensuremath{\textrm{err}_\textrm{max}}}

\newcommand\encircle[1]{%
  \tikz[baseline=(X.base)]
    \node (X) [draw, shape=circle, align=center, text width=1.2mm, inner sep=0] {\strut \small{#1}};}

\definecolor{lgray}{RGB}{250,250,250}

\SetKwComment{Comment}{$\triangleright$~}{}

\newtoggle{FALSE}
\togglefalse{FALSE}

\title{\retQSS: A Novel Methodology for Efficient Modeling and Simulation of Particle Systems in Reticulated Geometries}

\author{
  Lucio Santi\thanks{Corresponding author\newline\newline\copyright~2021. This manuscript version is made available under the CC-BY-NC-ND 4.0 license \url{http://creativecommons.org/licenses/by-nc-nd/4.0/}} \\
  Departamento de Computaci\'on\\
  FCEyN - UBA and ICC-CONICET\\
  Ciudad Universitaria, Pabell\'on 1\\
  C1428EGA, Buenos Aires, Argentina\\
  \texttt{lsanti@dc.uba.ar} \\
   \And
  Joaqu\'in Fern\'andez \\
  CIFASIS-CONICET \\
  Argentina\\
  \texttt{fernandez@cifasis-conicet.gov.ar}\\
  \And 
  Ernesto Kofman \\
  Departamento de Control \\
  FCEIA - UNR and CIFASIS-CONICET \\
  Argentina \\
  \texttt{kofman@cifasis-conicet.gov.ar}\\
  \And   
  Rodrigo Castro\\
  Departamento de Computaci\'on\\
  FCEyN - UBA and ICC-CONICET\\
  Ciudad Universitaria, Pabell\'on 1\\
  C1428EGA, Buenos Aires, Argentina\\
  \texttt{rcastro@dc.uba.ar} \\
}

\begin{document}

\maketitle

\begin{abstract}
This work presents retQSS, a novel methodology for efficient modeling and simulation of particle systems in reticulated meshed geometries. On the simulation side, retQSS profits from the discrete-event nature of Quantized State System (QSS) methods, which enable efficient particle tracking algorithms that are agnostic of the application domain. On the modeling side, retQSS relies on the standardized Modelica modeling language, yielding compact and elegant specifications of hybrid (continuous/discrete) dynamic systems. Combined together, these features offer a sound, general-purpose framework for modeling and simulation of particle systems. We show how the state-events that arise when particles interact with a reticulated mesh are seamlessly translated into easily tractable time-events. The latter can substantially  improve simulation performance in scenarios where discontinuities dominate the computational demand. We showcase the flexibility of our approach by addressing four case studies arising from different application domains. Performance studies revealed that retQSS can perform similarly to, and even outperform, well-known domain-specific particle simulation toolkits while offering a clear and sound accuracy control interface.
\end{abstract}


\section{Introduction}
\label{sec:intro}

Particle simulations have become an essential component in many modern disciplines such as computational fluid dynamics, high-energy physics, 3D rendering and agent-based modeling, to name a few. During the past decades, research and development efforts driven by varied needs (e.g., increasing demand for computing efficiency or better accuracy and stability of the methods) led to the emergence of diverse numerical methods and domain-specific software tools for particle simulation. Each application domain has thus adopted custom modeling methodologies which, in most cases, rely on general purpose programming languages not designed to address modeling problems.

In a broad sense, particle simulation can be classified according to how particles are transported. In the Lagrangian approach \cite{Zhang2007ComparisonSpaces}, the trajectories of a finite set of particles are individually tracked by solving ordinary differential equations (ODEs) governing their equations of motion. This approach is particularly attractive for capturing detailed spatio-temporal information of single particle trajectories. Also, it enables straightforward representations of dynamic characteristics such as external forces acting on a particle. On the other hand, the Eulerian framework \cite{Gui-rong2003SmoothedMethod} considers a control volume within which properties of interest of the underlying flow of particles are expressed as fields. Instead of identifying individual particles, it focuses on the activity at fixed points in space as time progresses.

This work presents \retQSS, a new software toolkit for efficient modeling and simulation of particle systems in three-dimensional spaces. In \retQSS, dynamic models track particles individually in the context of a reticulated geometry with which particles may interact as they are transported. Thus, \retQSS  follows a hybrid particle simulation approach in which particles are tracked in a Lagrangian setup while employing a background Eulerian setup for different purposes (e.g. interpolation of field properties). 

Instead of focusing on a single application domain, \retQSS pursues a more ambitious goal aimed at establishing a generic, yet rigorous, methodological framework for modeling and simulation of particle systems. An essential first step towards this goal is the adoption of QSS Solver \cite{Bergero2012QSSSolver2} as the underlying simulation engine. Models in QSS Solver are expressed in \uModelica, a high-level language for modeling dynamic systems, which is a subset of the more general Modelica language \cite{Fritzson1998ModelicaSimulation}. This mechanism is thus inherited in \retQSS, enabling succinct and elegant descriptions of particle models. Despite adopting QSS Solver as its primary simulation engine, other simulators may also profit from the particle simulation capabilities offered by \retQSS. Being simulator-agnostic by design, all of its functionality is exposed in a standalone C API wrapped into a Modelica package. This allows for straightforward connections with other Modelica-based simulators such as Dymola \cite{Bruck2002DymolaSimulation} or OpenModelica \cite{Fritzson2005TheEnvironment}.

A typical feature of particle simulations shared across different application domains is the need to track particles as they travel in a geometry, carefully identifying the crossing of boundaries between adjacent volumes. An important goal of \retQSS  is to provide efficient methods and algorithms to tackle this problem. To this end, \retQSS  leverages the optimized implementations of the Quantized State System (QSS) family of hybrid numerical methods \cite{Cellier2006,Kofman2001} offered by QSS Solver. These methods combine continuous with discrete-event dynamics to approximate continuous systems, exhibiting properties such as very efficient handling of discontinuities and dense output supported by polynomial approximations of state trajectories. Both properties are at the heart of the particle tracking algorithms implemented in \retQSS. Combined with the aforementioned high-level modeling capabilities, these efficient boundary crossing detection algorithms can treat computationally demanding \textit{state events} \cite{Cellier2006ContinuousSimulation} (the intersection between particle trajectories and volume boundaries) as easily tractable \textit{time events} (the time instants at which such intersections occur).

Despite being a general-purpose particle simulation methodology, \retQSS enables efficient modeling approaches to a wide range of application domains. As we shall see later in detail, \retQSS  can perform similarly to other well-known domain-specific particle simulation toolkits and, under certain circumstances, it can even achieve considerable performance gains. In particular, \retQSS is very efficient at simulating problems with intense boundary crossing activity. Other scenarios that may fully exploit \retQSS capabilities are those involving spatial dynamics that influence particle behavior.

Thus, we shall prove the flexibility of our approach by addressing four selected case studies of very different nature: bird flocking (an agent system with emergent behavior), a high-energy physics setup, a system of molecules interacting via an exponential potential (implementing the Molecular Dynamics method \cite{Rapaport2004TheSimulation}) and plasma flow (implementing custom variants of the Particle-In-Cell algorithm, widely used in plasma simulation \cite{Tskhakaya2007TheMethod}). We also carry out a comprehensive performance comparison against other related tools in order to assess the suitability of \retQSS as a plausible alternative for particle simulation in the corresponding application domains.

The rest of the paper is organized as follows. We start in Section \ref{sec:background} with an overview of the most relevant background concepts used throughout the manuscript (a summary of QSS theory, the QSS Solver simulation toolkit and the Modelica language). We continue in Section \ref{sec:motivation} with a description of the most essential concepts behind \retQSS, illustrated with a simple motivational example. A discussion about the high-level design and implementation details of our tool is then given in Section \ref{sec:design}. The motivational example introduced before is revisited again in Section \ref{sec:bouncing_balls_revisited}, this time with a complete description of a \retQSS model. In Section \ref{sec:case_studies} we show how \retQSS can be used to model particle systems of diverse nature, developing also thorough performance comparisons against related tools. Then, Section \ref{sec:related_work} puts \retQSS into context discussing related work in the field. Finally, Section \ref{sec:conclusions} provides a summary, conclusions and comments on our work in progress.

\section{Background}
\label{sec:background}

In this Section we present the essential concepts used along the article. We first discuss about QSS theory and continue with a brief summary of the standalone QSS Solver. Finally, we describe the high-level modeling language Modelica.

\subsection{Quantized State System (QSS) methods}
\label{sec:qss}

QSS methods replace the time discretization of classic numerical integration algorithms by the quantization of the state variables.  Given a time invariant ODE in its State Equation System (SES) representation,
\begin{equation}\label{eq:ODE_X}
 \mathbf{\dot x} = \mathbf{f} (\mathbf{x}(t) ,t)
\end{equation}

where $\mathbf{x}(t) \in \mathbb{R}^n$ is the state vector, the first order Quantized State System (QSS1) method \cite{Kofman2001} solves an approximate ODE called \textsl{Quantized State System}:
\begin{equation} \label{eq:ODE_q}
 \mathbf{\dot x} = \mathbf{f} (\mathbf{q}(t) ,t)
\end{equation}

\begin{figure}[h]
 \centering
 \includegraphics[scale=0.55]{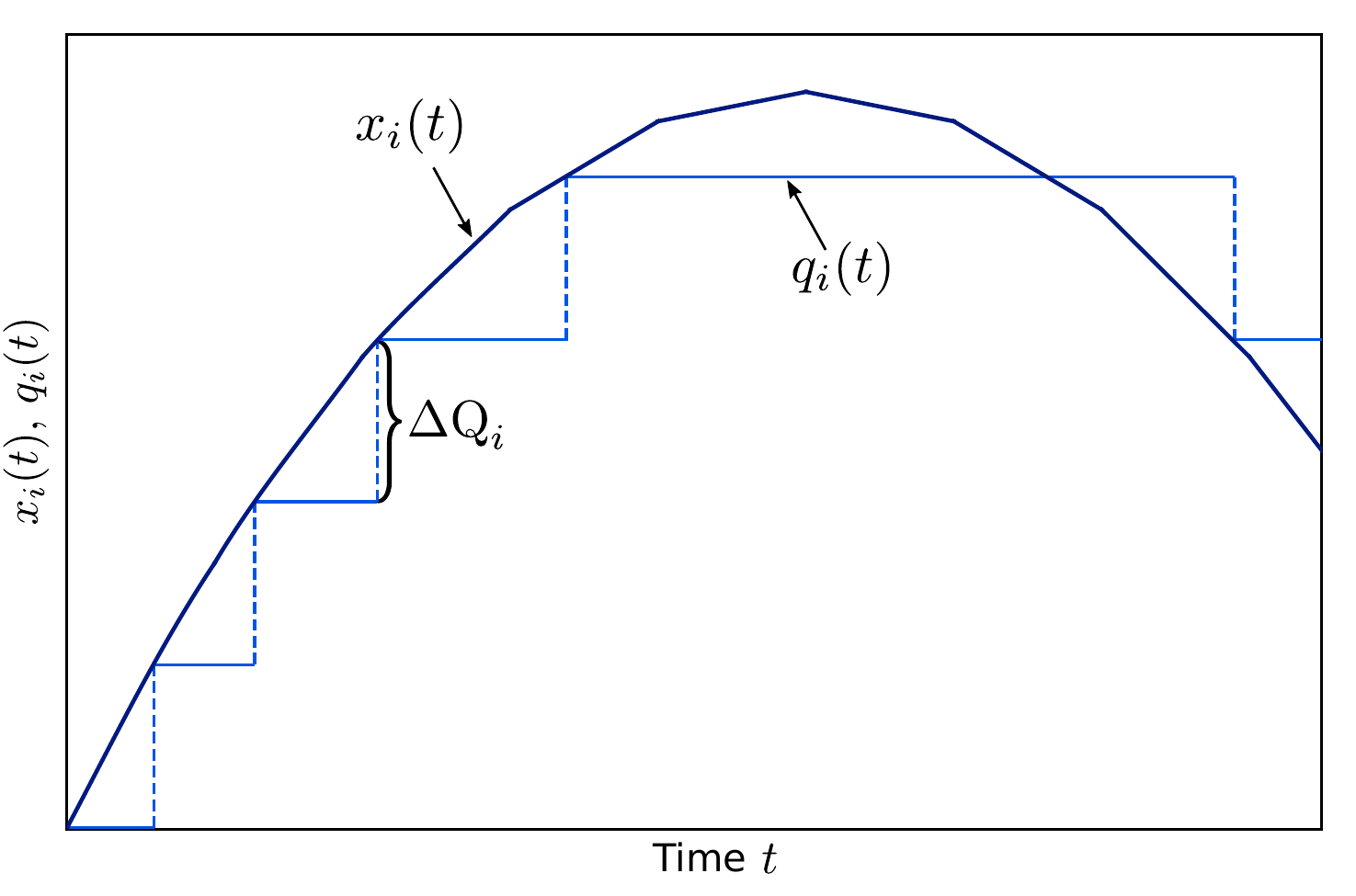}
\caption{QSS1 hysteretic quantization function}
\label{fig:hystquant}
\end{figure}

Here, $\mathbf{q}(t)$ is the \emph{quantized state} vector. Each component $q_i(t)$ follows a piecewise constant trajectory that only changes when its difference with the corresponding state $x_i(t)$ reaches the \textsl{quantum} $\deltaQ_i$. Denoting  $t_1, t_2, \ldots, t_k,\ldots $ the times at which the piecewise constant trajectory $q_i(t)$ changes, the quantized state trajectory is related to the corresponding state trajectory $x_i(t)$ as follows:
\begin{equation*}
 q_i(t) =
 \begin{cases}
  q_i(t_k) \quad &\text{if} \;|x_i(t)-q_i(t_k)| < \deltaQ_i	\\
  x_i(t) \quad &\text{otherwise}
 \end{cases}
\end{equation*}
for $t_k< t\leq t_{k+1}$, where $t_{k+1}$ is the first time after $t_k$ at which $|x_i(t)-q_i(t_k)| = \deltaQ_i$.  In addition, we consider that initially $\mathbf q(t_0)=\mathbf x(t_0)$. This defines an \emph{hysteretic quantization function} generating trajectories like those depicted in Figure~\ref{fig:hystquant}.

The quantum plays an equivalent role to that of the tolerance in variable step size algorithms. Here, the step size is usually controlled in order to fulfill a relative error tolerance and this can be also achieved in QSS by using a quantum that changes with the signal amplitude:
$$\deltaQ_i = \max\left(\dQRel \cdot |x_i|, \dQMin\right)$$
where $\dQRel$ is the \textsl{relative quantum} (relative tolerance) and $\dQMin$ is the \textsl{minimum quantum} (absolute tolerance).

Since the quantized state trajectories $q_i(t)$ are piecewise constant, then, provided that the system is autonomous (or that $\mathbf f(\cdot,t)$ is piecewise constant with $t$) the state derivatives $\dot x_i(t)$ also follow piecewise constant trajectories and, consequently, the states $x_i(t)$ follow piecewise linear trajectories. \blue{Thus, the numerical solution of Eq.~\eqref{eq:ODE_q} becomes straightforward and can be translated into a simple simulation algorithm. The corresponding pseudo-code is shown below in the listing for Algorithm \ref{alg:qss1} (we assume for simplicity that the system is autonomous).}

\SetInd{0.25em}{0.4em}
\DontPrintSemicolon
\begin{algorithm}
  \fontsize{10pt}{10pt}\selectfont
\caption{QSS1}
\label{alg:qss1}
\KwIn{$t_i$ (initial time), $t_f$ (final time), $\mathbf x_0$ (initial state), $\Delta\mathbf{Q}$ (quantum vector)}
\BlankLine
\Begin(Initialization){
$t \gets t_i$ \Comment*[r]{Initialize simulation time}
$\mathbf{x} \gets \mathbf{x_0}$ \Comment*[r]{Initialize continuous state vector}
$\mathbf{q} \gets \mathbf{x}$ \Comment*[r]{Initialize quantized state vector}

\ForEach{$j \in [1,n]$}{
  $\dot x_j \gets f_j(\mathbf{q}, t)$ \Comment*[r]{Compute $j$-th initial state derivative}
  $t^{\nxt}_j \gets t + \Delta Q_j/\abs{\dot x_j}$ \Comment*[r]{Time of next change in the $j$-th quantized state}
  $t^x_j \gets t$ \Comment*[r]{Time of last change in the $j$-th continuous state} 
  $t^q_j \gets t$ \Comment*[r]{Time of last change in the $j$-th quantized state}
}}
\BlankLine
\Begin(Simulation loop){
    \While{$t < t_f$}{\label{alg:qss_loop} 
        $t \gets \textrm{min}(t^{\nxt}_j)$ \Comment*[r]{Advance simulation time} \label{alg:qss_advance_time}
        $i \gets \textrm{argmin}(t^{\nxt}_j)$ \Comment*[r]{The $i$-th quantized state changes first}
      $e \gets t - t_i^x$ \Comment*[r]{Elapsed time since last $x_i$ update} 
      $x_i \gets x_i  + \dot x_i \cdot e$ \Comment*[r]{Update $i$-th state value}
      $q_i \gets x_i$ \Comment*[r]{Update $i$-th quantized state} \label{alg:qss_update_qi}
      $t^q_i \gets t$ \Comment*[r]{Time of last change in the $i$-th quantized state}
      $t^{\nxt}_i \gets t + \Delta Q_i/|\dot x_i|$ \Comment*[r]{Time of next change in the $i$-th quantized state}
      \ForEach{$j \in [1,n]$ \textrm{such that} $\dot x_j$ \textrm{depends on} $q_i$}{
        $e \gets t - t_j^x$ \Comment*[r]{Elapsed time since last $x_j$ update}
        $x_j \gets x_j  + \dot x_j  \cdot e$ \Comment*[r]{Update $j$-th state value}
        \lIf*{$j\neq i$}{
            $t_j^x \gets t$  \Comment*{Last $x_j$ update}
        }
        $\dot x_j \gets f_j(\mathbf{q}, t)$ \Comment*[r]{Recompute $j$-th state derivative}
        $t^{\nxt}_j \gets \textrm{min}(\tau > t)\, \textrm{subject to}\, \left| q_j - x_j(\tau) \right| = \Delta Q_j$ \Comment*[r]{Recompute time of next change in the $j$-th quantized state} \label{alg:qss_next_change_qj}
      }
      $t_i^x \gets t$ \Comment*[r]{Last $x_i$ update}
    }
}
\end{algorithm}

\blue{On each iteration of the main integration loop (line \ref{alg:qss_loop}), the first action is to determine the quantized state $q_i$ that changes first. Then the simulation time is advanced until the timestamp of that event (line \ref{alg:qss_advance_time}). The algorithm then advances the state value $x_i$ (using the fact that $\dot x_i$ is constant in this period) and computes the new quantized state $q_i=x_i$ (line \ref{alg:qss_update_qi}). This change in $q_i$ will affect some state derivatives $\dot x_j=f_j(\mathbf q,t)$ -- in particular, those in which $q_i$ explicitly appears in the expression of $f_j$. Thus, the algorithm recomputes the corresponding states $x_j$ and the next time of change for the associated quantized states $q_j$ (line \ref{alg:qss_next_change_qj}).} 

\blue{The time of the next change $t^{\nxt}_j$ is computed as the first time after $t$ at which the difference between the piecewise constant trajectory $q_j(t)$ and the piecewise linear trajectory $x_j(t)=x_j+\dot x_j \cdot (t-t^x_j)$ becomes equal to the quantum $\Delta Q_j$. This is,}
\blue{\begin{equation*}
 \abs{x_j+\dot x_j \cdot (t^{\nxt}_j-t^x_j)-q_j}=\Delta Q_j
\end{equation*}}
One of the drawbacks of the QSS1 method is that it performs a first-order approximation. Higher-order QSS methods (e.g. QSS2 and QSS3) follow the same basic principle as QSS1: in QSS$n$, $\mathbf{x}(t)$ follow piecewise $n$-th degree polynomial trajectories and $\mathbf{q}(t)$ follow piecewise $(n-1)$-th degree polynomial trajectories \cite{Kofman2001}.

\begin{figure}[h]
    \centering
    \includegraphics[scale=0.65]{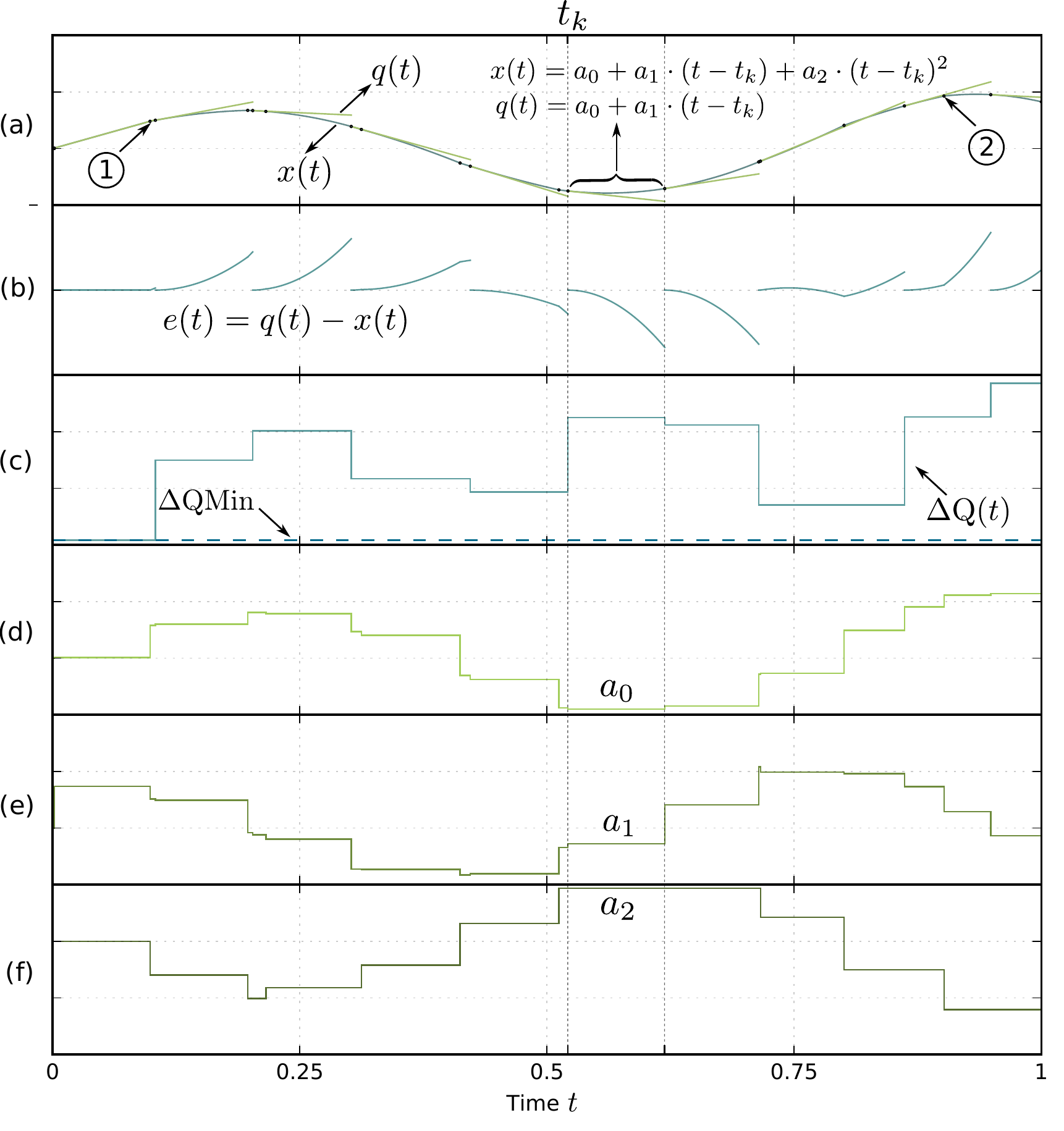} 
    \caption{Illustrative example of a QSS2-based simulation and its main underlying QSS concepts}
    \label{fig_qssplot}
\end{figure}

In order to put all these concepts together, Figure \ref{fig_qssplot} presents a QSS2 simulation of the position in the $\hat{x}$ axis of a charged particle in a constant magnetic field (the model in use is the ODE system in Equation~\eqref{eq:lorentz}, which will be revisited in Section \ref{sec:hep_model}). Figure \ref{fig_qssplot}a shows the solution state variable $x(t)$ and its corresponding quantized state variable $q(t)$, which follow piecewise quadratic and linear trajectories, respectively. Each dot in the curve marks endings and the commencements of adjacent polynomial sections. Sections such as \small{\encircle{1}} affect the coefficients of the state variable $x(t)$. They happen due to a reaction to an update originated from another state variable that affects the state derivative $\dot{x}(t)$. On the other hand, sections such as \small{\encircle{2}} happen when the quantum $\deltaQ$ (the maximum deviation allowed between $q(t)$ and $x(t)$) is reached. In these situations, the coefficients of $q(t)$ are recomputed by quantizing the state variable $x(t)$.

The difference between $q(t)$ and $x(t)$ is the error $e(t)$ incurred by the method and is shown in Figure \ref{fig_qssplot}b. It gets determined by the user-supplied accuracy-related parameters $\dQRel$ and $\dQMin$, also shown in Figure \ref{fig_qssplot}c. If we consider for example the section starting at time $t_k$, we can see that both $q(t)$ and $x(t)$ evolve until the difference between them reaches $\deltaQ$. At that time, $q(t)$ is updated by quantizing $x(t)$, giving rise to a new polynomial section for $q(t)$ in the plot. This change is propagated to the ODE system by evaluating those state variables whose right-hand side depends on this variable. The coefficients of the polynomial approximations of $q(t)$ and $x(t)$ are presented in Figures \ref{fig_qssplot}d, \ref{fig_qssplot}e and \ref{fig_qssplot}f.

\subsubsection{Properties of QSS methods}
\label{sec:qss_properties}

In QSS, computational steps are produced when a quantized variable $q_i$ changes after reaching a difference with $x_i$ that exceeds the quantum $\deltaQ_i$. This change is propagated to those state derivatives that depend on $x_i$. In consequence, each step involves changes in one quantized variable and in a subset of the state derivatives. QSS inherently exploits this fact performing computations only when and where these changes occur. This is particularly relevant in the context of large, sparse systems that experience activity in a few states while the rest of the system remains inactive \cite{Grinblat2012QuantizedNetworks}.

Another important feature of the QSS methods is that they are very efficient at simulating systems with frequent discontinuities \cite{Grinblat2012QuantizedNetworks, Migoni2015Quantization-basedSupplies, Bergero18Building}. These are modeled by zero-crossing functions expressed in terms of the QSS polynomials. Hence, detecting a discontinuity calls only for finding the roots of a polynomial, which is computationally inexpensive for at most third-order QSS methods. Once a discontinuity is detected, the algorithm handles it as an ordinary step, as each step is in fact a discontinuity in a quantized variable. Thus, the occurrence of a discontinuity implies only some local calculations to recompute the state derivatives that are directly affected by that event. The simulation does not need to be restarted as it is typically required in classic integration algorithms.

QSS1 to QSS3 provide global error bound properties. These establish that, when simulating stable linear time invariant (LTI) systems, the numerical solution differs from the analytical solution in a quantity that is linearly bounded with the quantum $\deltaQ$ \cite{Cellier2006ContinuousSimulation}. Thus, in order to increase the accuracy by a given factor $f$, the quantum should be reduced by the same factor. In QSS$n$, the number of extra calculations required to achieve this outcome is proportional to the $n$-th root of $f$. For example, if $f = 10^{6}$, in QSS1 we would need to multiply by $10^6$ the number of calculations. In QSS2, we would need to multiply them by $\sqrt{10^6} = 1000$, and in QSS3, by $\sqrt[3]{10^6} = 100$ \cite{Kofman2006QSS3}.

\subsection{The Standalone QSS Solver toolkit}
\label{sec:solver}

Most implementations of QSS methods are provided by general-purpose discrete event simulation engines such as PowerDEVS \cite{Bergero2010PowerDEVS:Simulation}. This generality usually brings about unnecessary CPU overheads (due to the underlying message-passing and/or event scheduling mechanisms) when the systems to be simulated are primarily continuous. The standalone QSS Solver, on the other hand, is an open-source software that offers optimized implementations of the whole family of QSS methods, improving in more than one order of magnitude the computation times of previous discrete event implementations \cite{fernandez2014stand}. QSS Solver simulates models that can contain discontinuities. These models are represented as follows:
\begin{equation} \label{eq:qssgen}
   \mathbf{\dot x}(t)=\mathbf f(\mathbf x,\mathbf d,t)
\end{equation}
where 
 $\mathbf d$ is a vector of discrete variables that can only change when a condition 
\begin{equation}\label{eq:zc}
 \ZC_i(\mathbf x,\mathbf d,t)=0
\end{equation}
is met (for some $i\in \{1,\dots,z\}$). The components $\ZC_i$ form a vector of \textsl{zero--crossing functions} $\mathbf{ZC}(\mathbf x,\mathbf d,t)$. When such a zero--crossing condition is verified, the state and discrete variables can change according to the corresponding \textsl{event handler}:
\begin{equation}\label{eq:evh}
(\mathbf x(t),\mathbf d(t)) = \Handler_i(\mathbf x(t^-),\mathbf d(t^-),t)
\end{equation}  

These models are simulated using QSS methods that approximate Equation~\eqref{eq:qssgen} by
\begin{equation} \label{eq:qssd}
   \mathbf{\dot x}(t)=\mathbf f(\mathbf q,\mathbf d,t)
\end{equation}
where each component $q_i(t)$ is a piecewise polynomial approximation of the corresponding component of the state $x_i(t)$, as explained in Section \ref{sec:qss}. 

The simulation is performed by three modules interacting at runtime:

\begin{enumerate}
\item The \textbf{Integrator}, that integrates Equation~\eqref{eq:qssd} assuming that the piecewise polynomial quantized state trajectory $\mathbf q(t)$ is known. It is in charge of advancing the simulation time executing the simulation steps. Each of these may correspond to a change in a quantized variable $q_i$ or to the execution of an event handler $\Handler_i$ triggered by a zero--crossing condition $\ZC_i(t)=0$.

\item The \textbf{Quantizer}, that computes $\mathbf{q}(t)$ from $\mathbf{x}(t)$ according to the QSS method in use and their tolerance settings (there is a different Quantizer for each QSS method). That way, it provides the polynomial coefficients of each quantized state $q_i(t)$ and computes the next time at which a new polynomial section starts (i.e., when the condition $|q_i(t)-x_i(t)| = \deltaQ_i$ is met).

\item The \textbf{Model}, which computes the scalar state derivatives $\dot x_i=f_i(\mathbf q, \mathbf d, t)$, the zero--crossing functions $\ZC_i(\mathbf x,\mathbf d,t)$, and the corresponding event handlers $\Handler_i(\mathbf q,\mathbf d,t)$. Besides, it provides the structural information required by the algorithms (represented as a set of four binary incidence matrices).
\end{enumerate}

\begin{figure}[h]
    \includegraphics[width=10cm]{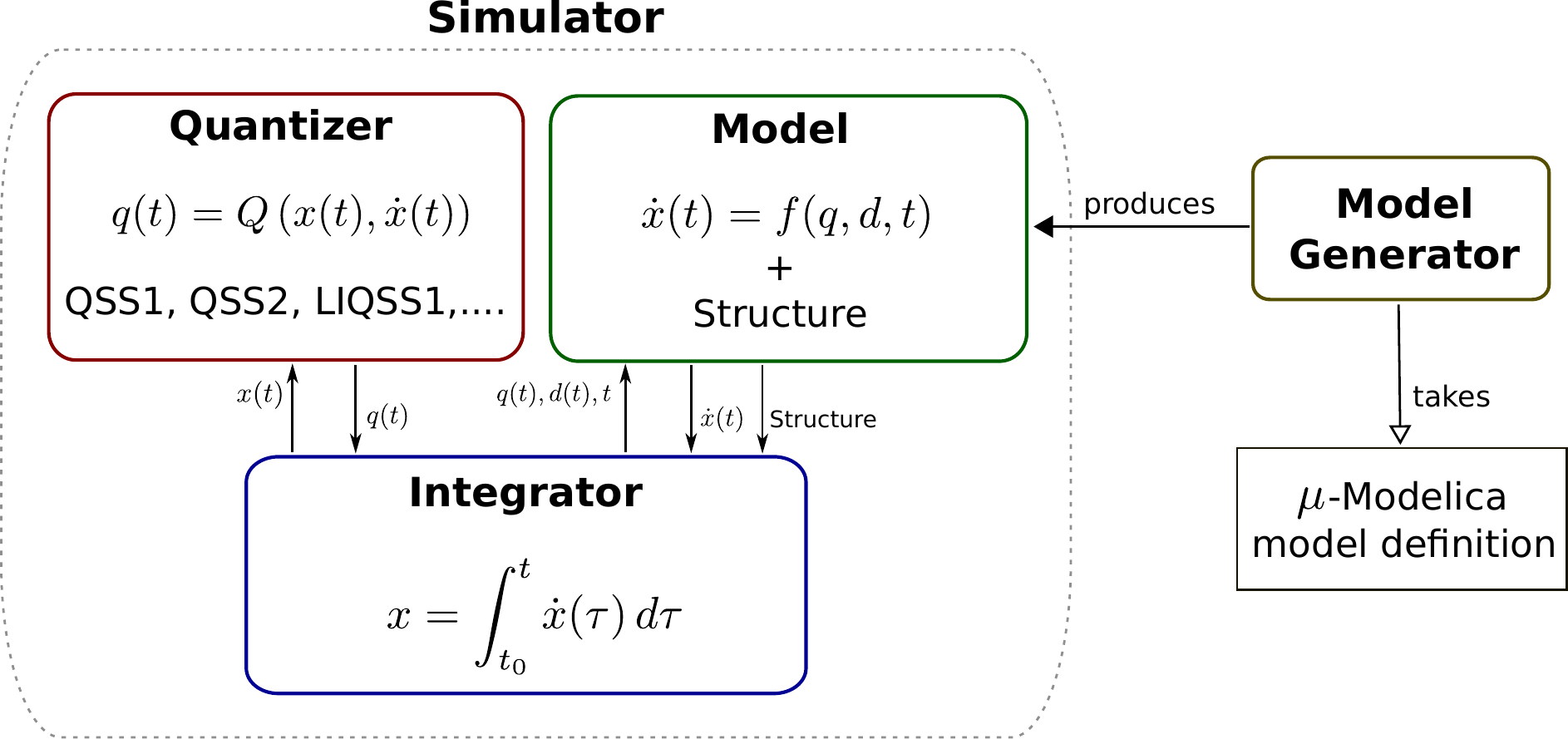}
  \caption{Basic interaction scheme of QSS Solver modules}
  \label{fig:basicscheme}
\end{figure}

The structural information of the Model is automatically extracted at compile time by a \textbf{Model Generator} module. This module takes a standard model described in $\mu$-Modelica \cite{Bergero2012QSSSolver2}, a subset of the more general Modelica modeling language (introduced next in Section \ref{sec:modelica}), and produces an instance of the Model module as required by the QSS solver. This instantiation includes the structural information and the possibility of separately evaluating scalar state derivatives. Figure \ref{fig:basicscheme} shows the basic interaction scheme between the three modules mentioned above. A simple graphical user interface integrates the solver engine with the modeling front-end and plotting and debug ancillary tools.

\subsection{The Modelica modeling language}
\label{sec:modelica}

In an effort to unify the different modeling languages used by the different modeling and simulation tools, a consortium of software companies and research groups proposed an open, unified object-oriented modeling language called \emph{Modelica} \cite{Fritzson1998ModelicaSimulation, Fritzson2010Principles2.1}. 

Modelica allows the representation of continuous time, discrete time, discrete event and hybrid systems. Elementary Modelica models are described by sets of differential and algebraic equations that can be combined with algorithms specifying discrete evolutions. These elementary models can be connected to other models to assemble complex model structures, facilitating the construction of multi--domain models.

Modelica models can be built and simulated using different software tools. OpenModelica \cite{Fritzson2005TheEnvironment} is the most  complete open source package, while Dymola \cite{Bruck2002DymolaSimulation} and Wolfram System Modeler are the most used commercial tools. QSS Solver has also the capability of simulating Modelica models provided that they are previously flattened and the resulting systems of differential algebraic equations are sorted and converted into sets of ODEs \cite{Bergero2012QSSSolver2}.

\section{Key concepts and motivation}
\label{sec:motivation}

The guiding principle of \retQSS is to offer a novel methodology for modeling and simulation of particle systems in meshed geometries. In a broad sense, a \textsl{particle} is a physical entity moving in a three-dimensional space, possibly interacting with its environment throughout its trajectory. These interactions may have diverse effects, altering not only the behavior of the particle but also certain properties of the geometry. Thus, \retQSS models are characterized by integrating continuous dynamics given by particle trajectories with discrete dynamics arising from the interactions between particles and geometry. From a modeling perspective, this integration happens transparently as a consequence of the use of a high-level modeling language such as \uModelica, allowing for a decoupling of the modeling logic and its concrete implementation. Another visible effect of this decoupling is a straightforward geometry management. \retQSS provides access to different geometrical objects (volumes, faces and vertices) while hiding the underlying implementation details. This way, the same modeling logic can usually operate with different geometries, avoiding the need to update portions of the model code after changing its input. The most relevant geometry format adopted by \retQSS is text-based VTK, which is widely used by several other related tools such as ParaView \cite{Ayachit2015TheApplication}, an open source application for scientific visualization, and OpenFOAM, a toolkit for computational fluid dynamics \cite{Jasak2007OpenFOAM:Simulations}. 

One of the most attractive features of \retQSS, derived from its primary design goal, is its great versatility to tackle a wide pallette of problems. As we shall see later, \retQSS can effectively model agent systems with emergent behavior (e.g., bird flocking), subatomic particles in high-energy physics setups, plasma flow and systems of molecules with intermolecular interactions given by custom force fields.

\subsection{A simple example: bouncing balls with obstacles}
\label{sec:bouncing_balls}

\begin{figure}[h]
    \centering
    \includegraphics[scale=0.25]{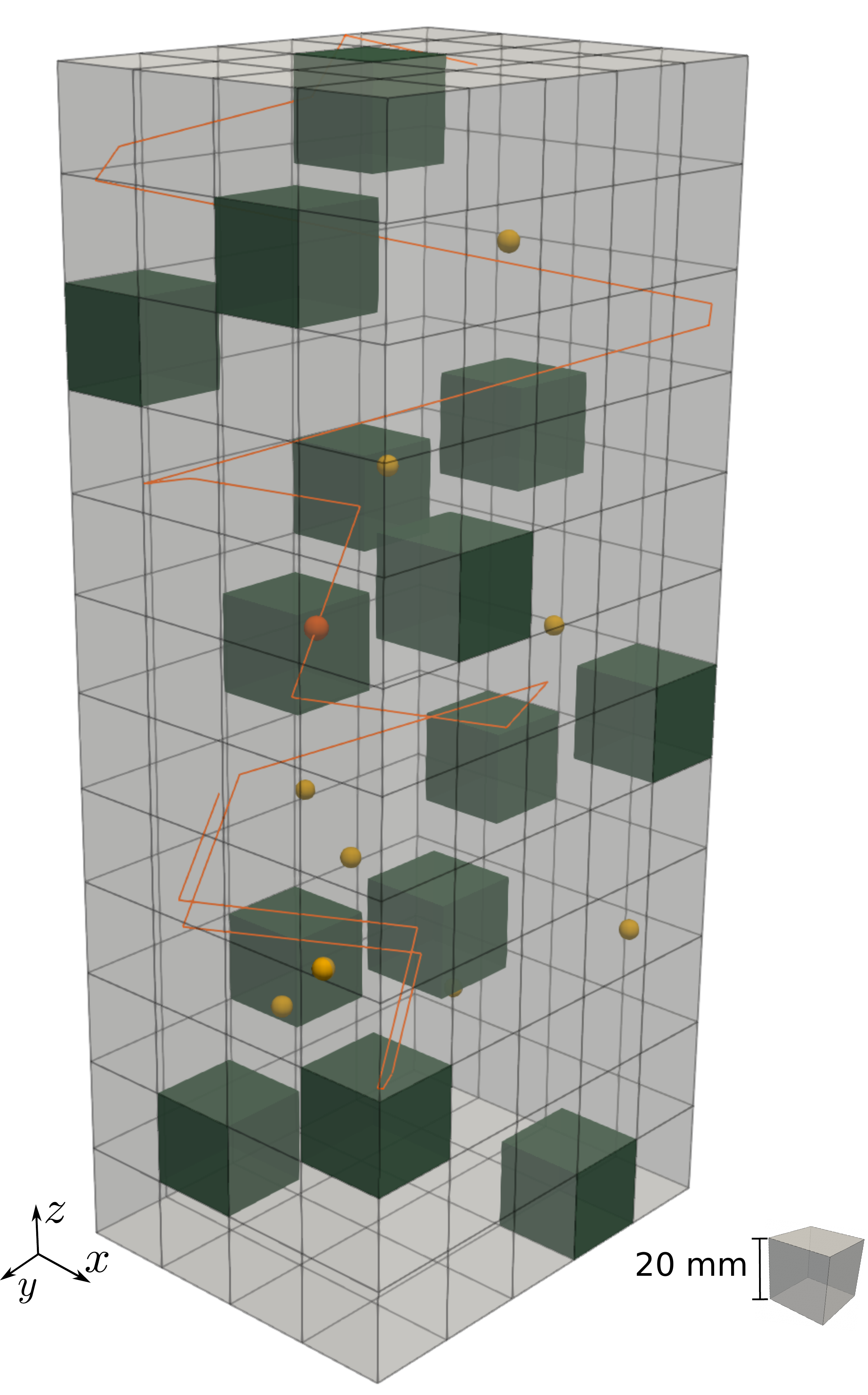}\par 
    \caption{Balls bouncing in a cube mesh with obstacles}
    \label{fig_bouncing_balls}
\end{figure}

We shall now discuss how the structure of a \retQSS model materializes the concepts outlined above. To this end, suppose we are interested in modeling a collection of balls bouncing within a chamber containing some obstacles. We will assume that the balls do not interact directly with each other. Figure \ref{fig_bouncing_balls} shows ten balls in a chamber represented by a lattice of cubes where obstacles are modeled as a subset of the cubes. 

The following considerations would need to be addressed when modeling such a problem:
\begin{enumerate}
    \item\label{balls_trajectories} How ball trajectories should be defined (e.g. as a system of ODEs to be integrated by a numerical solver),
    \item\label{balls_geometry} How to represent the chamber as a 3D geometry, including a proper characterization of the obstacles, and
    \item\label{balls_interaction} How to implement the interactions between balls and the geometry (i.e., deciding when and how a ball should bounce).
\end{enumerate}

By adopting a simulation engine such as QSS Solver, \retQSS inherits some elegant and efficient mechanisms to tackle these issues. On the one hand, \retQSS profits from its optimized, state-of-the-art implementations of QSS methods (as we shall see later, this enables efficient simulations in several application domains). Also, models are written in \uModelica, a high-level modeling language where continuous systems are described in a mathematical form  (e.g., those arising from the dynamics of the balls). Through \uModelica~ it is also possible to provide a compact and declarative algorithmic description of the interactions between balls and geometry in the form of discrete events. The particle tracking features of \retQSS allow for capturing the time instants at which the balls hit any obstacle, being therefore straightforward to define such discrete events.

In \retQSS models, different  geometries can be transparently incorporated by supplying a geometry file to the initialization algorithm. Thus, experimenting with different geometries is usually as simple as changing the input geometry file. In the case of the example at hand this implies that the exact same model can be used to test how balls bounce within chambers and against obstacles of very different shapes (provided suitable geometry descriptions are available in any of the supported file formats).

Obstacles may be modeled as a subset of the cells (or volumes) conforming the geometry. For this purpose, the modeler chooses which of the volumes are marked as obstacles. Volumes are exposed by \retQSS with an associated unique identifier so that models can operate explicitly with them. One of the most important modeling tasks is the definition and assignment of dynamic, piecewise constant properties to volumes and particles. Thus, one possible modeling choice is to define  an \textit{obstacle} boolean property, and set it \textit{true} only for selected volumes. Along the same lines, the concept of dynamic properties can be applied to each ball, or more generally speaking, to each particle in the system. This would allow for modeling different types of interactions between different particles and different volumes, providing a very flexible setting to express varied types of dynamics.

We shall revisit this motivating bouncing balls example and analyze it in more detail in Section \ref{sec:bouncing_balls_revisited}, after having delved into the technical underpinnings of \retQSS.

\section{Design and implementation}
\label{sec:design}

\begin{figure}[h]
    \centering
    \includegraphics[scale=0.6]{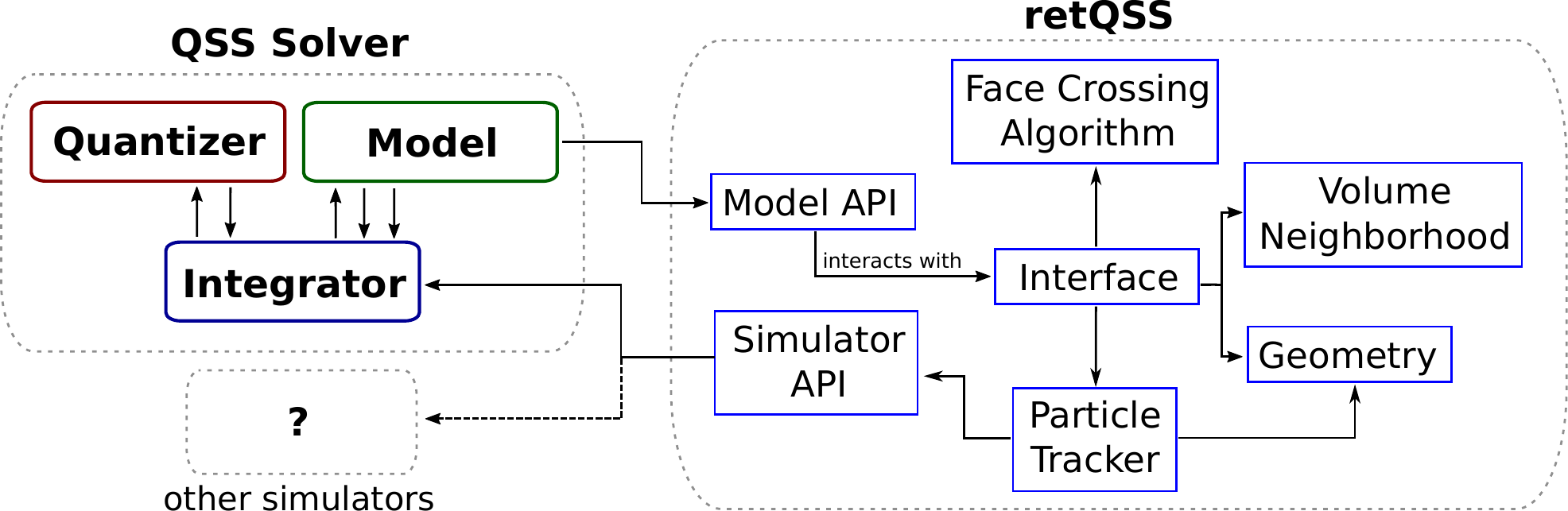}\par 
    \caption{High-level software architecture of \retQSS}
    \label{fig_arch}
\end{figure}

From a software engineering perspective, \retQSS is an object system that follows some  principles and best practices of object oriented software design, in particular encapsulation, loose coupling and high cohesion \cite{Zou2002MigrationApproach}. To this end, each object is responsible for a well-defined set of cohesive tasks. As shown in Figure \ref{fig_arch}, the entry point to \retQSS is the \textsl{Model API}, which exposes the full set of available queries in the form of C functions. Queries are grouped in eight categories, which will be detailed next in Section \ref{sec:model_api}. 

Upon receiving a query request, the model API forwards it to an internal \textsl{Interface} in charge of orchestrating and managing interactions between the different objects of the \retQSS ecosystem. One of such objects is the \textsl{Geometry}, which mediates access to other related objects (volumes, faces and vertices) and implements general-purpose geometry queries and algorithms. The task of tracking particles as they travel across the geometry is assigned to the \textsl{Particle Tracker}. It stores tracking information (e.g., current and last volumes traversed, entry and exit faces and next scheduled exit time) that is publicly exposed through the model API as well as consumed by other objects. A central aspect of particle tracking is the detection of the time instants at which a particle crosses a volume boundary. The \textsl{Face Crossing Algorithm} implements an efficient strategy to achieve this, leveraging the QSS polynomial approximations of particle trajectories and the faces of the polyhedrons that tessellate the geometry. 

An important feature of \retQSS is the concept of \textsl{neighborhoods}. Particle neighborhoods are based upon volume neighborhoods, which can be computed by several different algorithms that can be chosen even during simulation time. The \textsl{Volume Neighborhood} object encapsulates these strategies and exposes a common, transparent interface to operate with volume neighborhoods. 

Despite having adopted QSS Solver as the primary simulation engine, \retQSS is designed to be simulator-agnostic. On the one hand, \retQSS is compiled as a standalone library that is later linked into the model executables produced by QSS Solver. Also, simulator-dependent behavior was abstracted away in an interface to perform backward queries to the simulation engine (e.g., providing access to values of state variables or coefficients of the QSS polynomials). This functionality was encapsulated in an ad-hoc class hierarchy. The \textsl{Simulator API} object is an instance of a concrete member in this class hierarchy. As a consequence, we expect to lower the software design burden required to make \retQSS work with simulation engines other than QSS Solver. \blue{\retQSS is an open-source software project and its source code is available at \cite{retQSSRepo} (it includes a copy of the QSS Solver toolkit).}

In what follows we shall provide a comprehensive description of \retQSS, starting with the supported geometry formats and the chosen computational geometry library. We then introduce the important concept of \textsl{first-class objects}: those that have public visibility and with which the model can operate with by supplying unique IDs. Then, we discuss the supported types of volume neighborhoods and we define the related notion of \textsl{particle neighborhood}, which is at the heart of some of the selected case studies we will cover in Section \ref{sec:case_studies}. After this, we explain in detail the boundary crossing detection algorithm. Finally, we describe the model API and summarize its most relevant queries.

\subsection{Geometry definition and management}
\label{sec:geometry}

\retQSS relies upon the Computational Geometry Algorithms Library (CGAL) \cite{Fabri2009CGAL:Library} to implement most of its underlying geometrical algorithms and internal data structures. CGAL is an open source C++ software library that offers a broad collection of computational geometry resources. In particular, \retQSS focuses mostly on its 3D fast intersection and distance computation features such as the AABB tree data structure and its related algorithms. Each volume in the input meshes is internally represented by a 3D surface mesh that specializes the \texttt{CGAL::Surface\_mesh} class, along with an AABB tree that uses its polygonal faces as primitives. These data structures are assembled during the bootstrapping phase of \retQSS, when the geometry input file is parsed. The AABB tree serves multiple purposes, e.g. determining whether a point lies outside or inside the volume or computing intersections of rays or segments with one of the volume's faces. The boundary crossing detection algorithm, covered in Section \ref{sec:crossing_algorithm}, employs the latter when finding intersections with volumes featuring a large number of faces.

\paragraph{Supported file formats}
\label{sec:formats}

Geometry can be specified in two file formats: the Visualization Toolkit (VTK) \cite{Schroeder2006TheGraphics} format (in particular, the ASCII-based mode for unstructured grids) and the Object File Format (OFF) \cite{Rost1989OFFFormat}. The latter is the default format supported by CGAL and can be easily used to represent arbitrary convex polyhedrons with any number of faces, whereas the former is more convenient for compact representations of meshed domains. In fact, we typically use VTK files since they are also supported by many other related tools, as discussed in Section \ref{sec:motivation}.

\subsection{First-class objects}
\label{sec:objects}

\begin{figure}[H]
    \centering
    \includegraphics[scale=0.13]{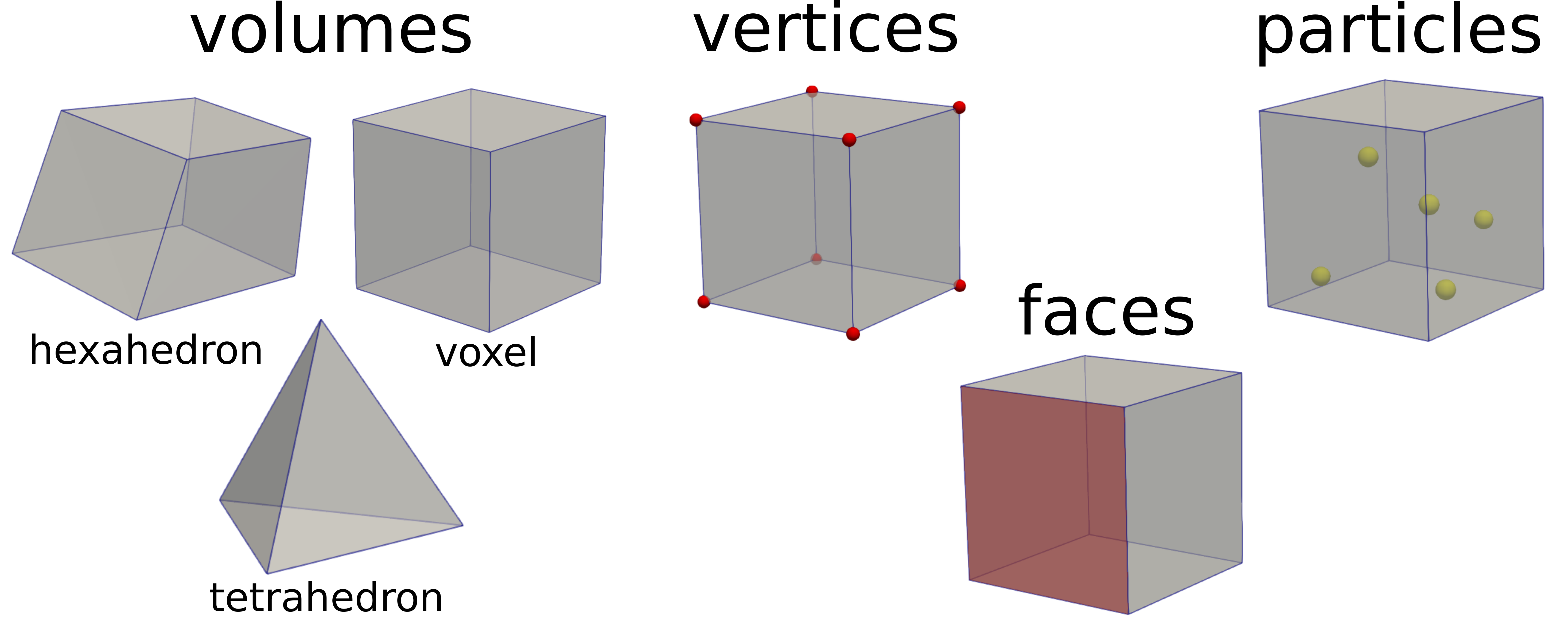}\par 
    \caption{First-class objects}
    \label{fig_objects}
\end{figure}

Models can explicitly operate and interact with \textsl{first-class objects}, a special kind of \retQSS entities that have public visibility and an associated unique identifier. A common feature of first-class objects is that they can store \textsl{properties}: an association between a character key and a numerical (or vectorial) value that can be set, queried or updated at any point during the simulation. Figure \ref{fig_objects} shows the available first-class objects in \retQSS: \textsl{volumes}, \textsl{vertices}, \textsl{faces} and \textsl{particles}.

Volume objects in this Figure are \textsl{polyhedral volumes}, i.e., they are defined by an underlying convex polyhedron. For VTK-based geometries, \retQSS supports so far hexahedral, tetrahedral and voxel cells, as the Figure shows (arbitrary convex polyhedrons can be supplied via OFF files). There is however another type of volume, the \textsl{open volume}, which is used to represent the exterior of the geometry. \retQSS automatically instantiates one such volume and gives it the ID 0. This is an essential feature so that particles can leave and enter the geometry, which is specially relevant e.g. for rigid geometry boundaries that should make the particles bounce back to the interior.

As we shall see, object properties have interesting and diverse applications. For example, in plasma models (Section \ref{sec:plasma}), vertex properties store aggregated particle charge and electric potential of every node in the domain grid, whereas bird flocking models (Section \ref{sec:boids}) can leverage this feature to model the presence of wind in different parts of the geometry. 

\subsection{Volume and particle neighborhoods}
\label{sec:neighborhoods}

\begin{figure}[hb!]
    \hspace{-1.5cm}
    \begin{tabular}{cc}
      \subfloat[Radial volume neighborhood\label{fig_radial}]
        {
            \includegraphics[scale=0.3,valign=t]{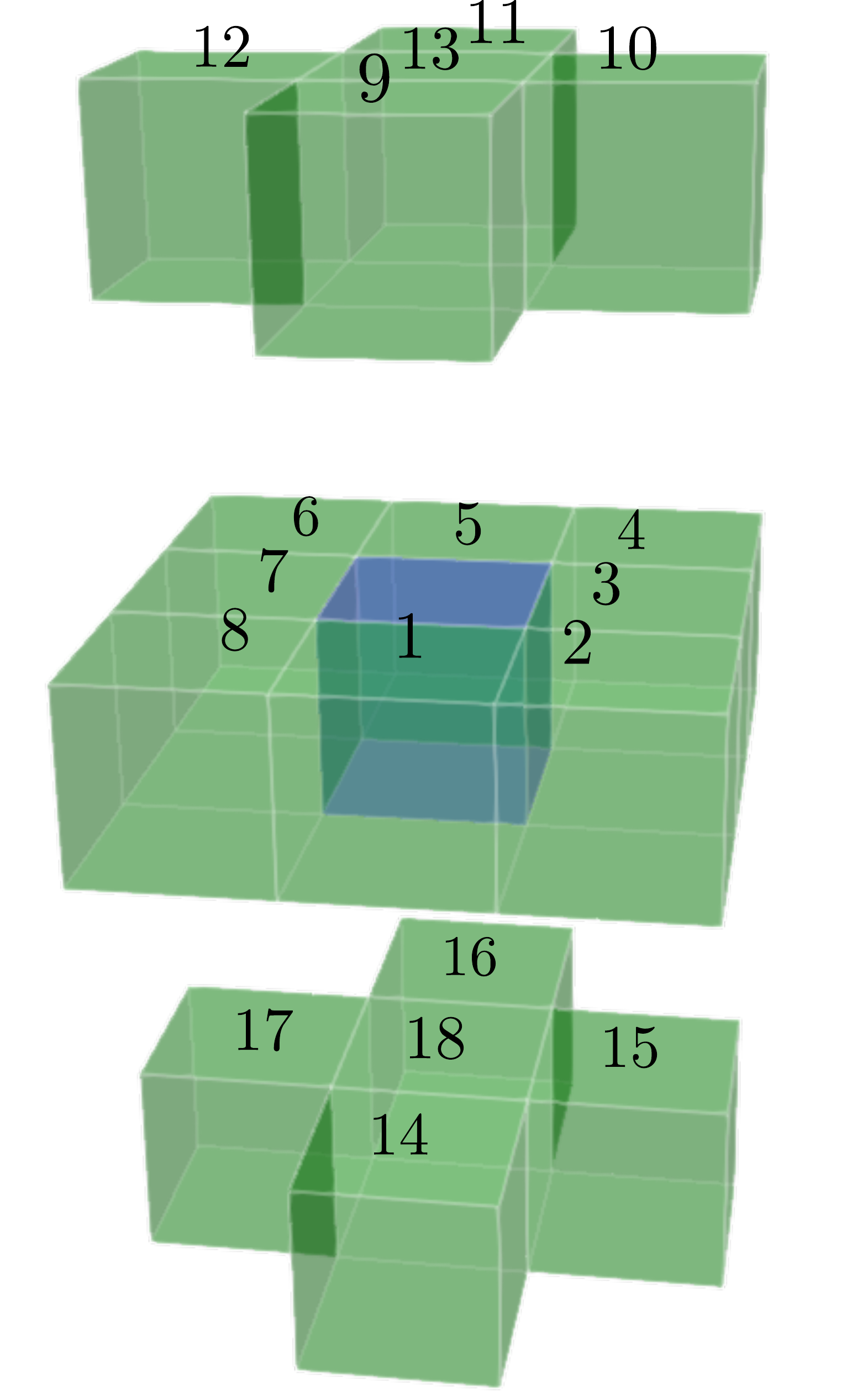}
        }   &
      \subfloat[Periodic radial volume neighborhood supporting a particle neighborhood\label{fig_periodic}]
        {
            \includegraphics[scale=0.38,valign=t]{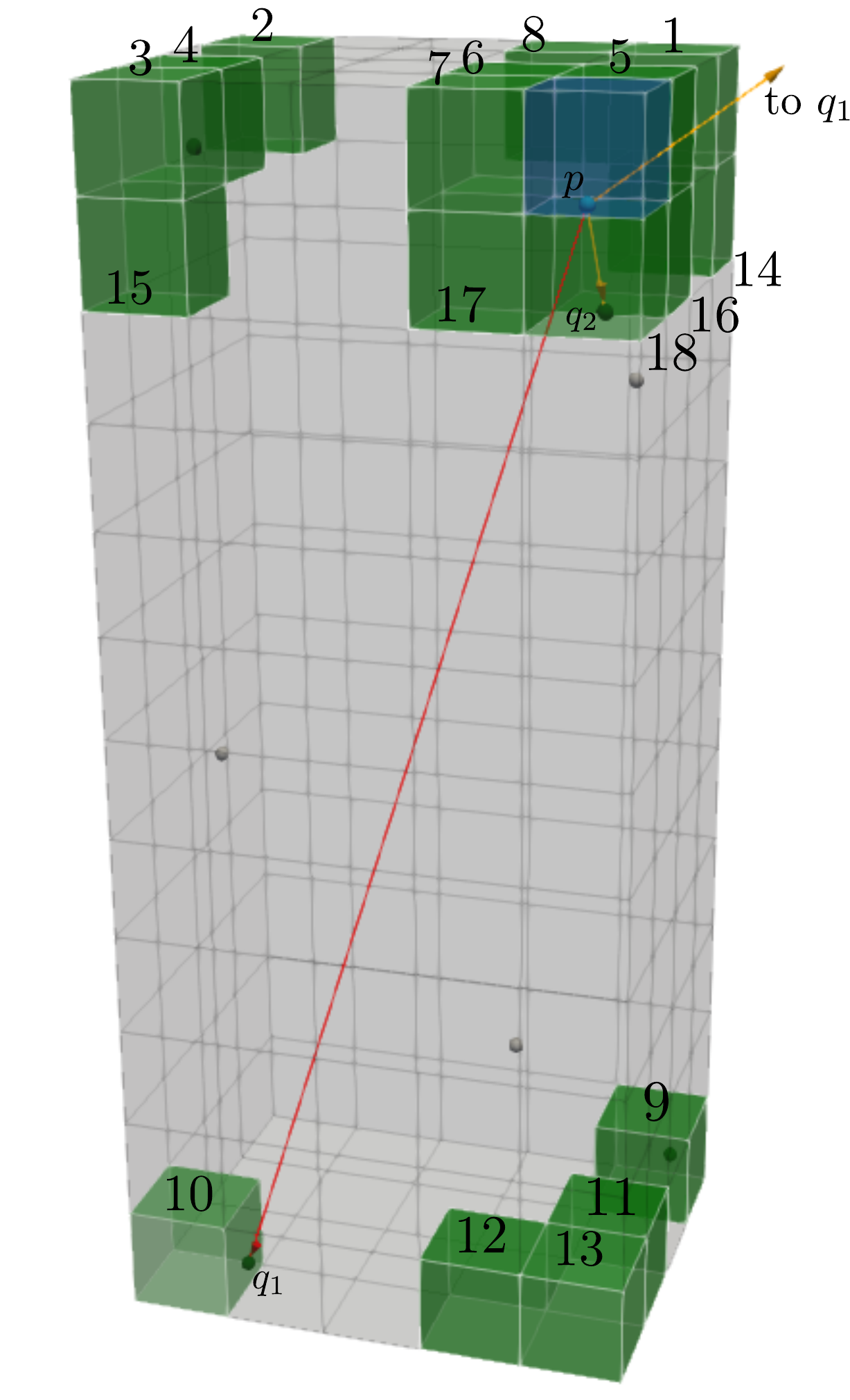}         \vphantom{\includegraphics[scale=0.38,valign=t]{images/7b-Periodic_Volume_Neigh.pdf}  }
        }    
    \caption{Illustrative example of volume and particle neighborhoods}
    \label{fig_neighborhoods}
    \end{tabular}
\end{figure}

Each volume has a \textsl{neighborhood}: a (possibly empty) set of volumes that can be calculated following different strategies:
\begin{itemize}
    \item A \textsl{face-sharing} approach, the default, in which a volume $u$ is considered neighbor of volume $v$ iff they both share a common face,
    
    \item A \textsl{vertex-sharing} algorithm in which a volume $u$ is considered neighbor of volume $v$ iff they both share at least one common vertex,
    
    \item A \textsl{radial} algorithm in which a volume $u$ is considered neighbor of volume $v$ iff the distance from $v$'s centroid to $u$ is no greater that a fixed value $r$ supplied as argument,
    
    \item A generalization of the previous strategy, the \textsl{periodic radial} neighborhood, which assumes periodic geometry boundaries (both strategies are illustrated in Figure \ref{fig_neighborhoods}), and
    
    \item A \textsl{file-based} approach, in which a text file specifies, for each volume in the geometry, the IDs of its neighbors.
\end{itemize}

During the simulation, models may switch freely between these algorithms. For efficiency purposes, neighborhoods are computed on-demand (i.e., neighborhood of volume $v$ is computed only when a \retQSS query requires access to $v$'s neighbors) and cached for future reuse. This cache is flushed every time the underlying algorithm is changed. \retQSS also offers the possibility of precomputing all volume neighborhoods up-front. 

Volume neighborhoods are particularly important as they lay the foundations for \textsl{particle neighborhoods}, an essential concept behind several \retQSS models (two such examples shall be explored in Section \ref{sec:case_studies}). In a similar fashion, the neighborhood of particle $p$ is a (possibly empty) set of particles: those inside any of the neighboring volumes where $p$ is located.

In Figure \ref{fig_neighborhoods} we can see how these concepts interplay with each other. We have a cube mesh geometry where a source volume (in blue) has its neighbors (in green) computed with the radial algorithm (Figure \ref{fig_radial}) and the periodic radial algorithm (\ref{fig_periodic}). Since the source is a boundary volume in \ref{fig_periodic}, we see that neighbor 1 appears on the opposite side of the geometry. In the case of e.g. neighbor 10, we need two jumps from the source to locate it in \ref{fig_radial} (up and right). Translating these jumps to \ref{fig_periodic}, we face twice the boundary periodicity, thus making opposite moves both times.

Figure \ref{fig_periodic} also shows a particle $p$ inside the source volume. It has four neighbors (green particles), two of which are highlighted: $q_1$ and $q_2$, inside volumes 10 and 18, respectively. Two vectors join $p$ with these particles. These are the \textsl{Euclidean vectors} whose norms are the Euclidean distances between $p$ and $q_i$. We also distinguish another vector sourced at $p$ that escapes the geometry: this is the \textsl{shortest vector} to $q_1$, which differs from the Euclidean vector as $q_1$'s volume is reached through periodic boundaries.

\retQSS offers special queries devoted to particle neighborhood traversal that facilitate access to such Euclidean and shortest vectors. This will be further explained in Section \ref{sec:model_api}.

\subsection{Boundary crossing detection algorithm}
\label{sec:crossing_algorithm}

\begin{figure}[h]
    \centering
    \includegraphics[scale=0.65]{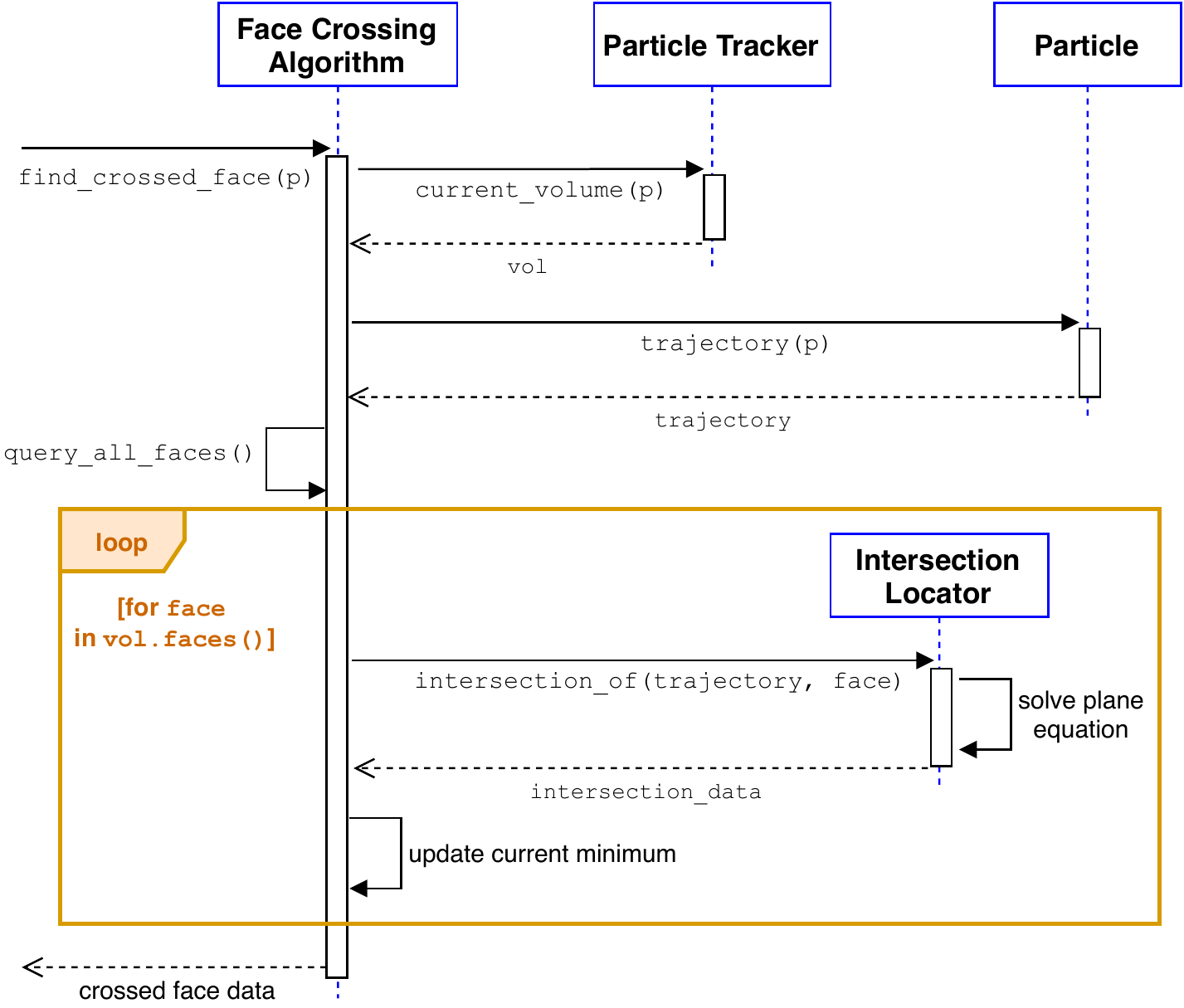}\par 
    \caption{Boundary crossing detection algorithm}
    \label{fig_crossing_algorithm}
\end{figure}

A particle traveling in a meshed geometry may visit several different volumes. When moving inside any such volume, it can eventually cross one of the volume's  faces to enter into an adjacent volume. In order to properly track particles along the simulation, it is thus important to identify the time instants of these  boundary crossings. The boundary crossing detection algorithm is a central aspect of \retQSS that solves this problem efficiently by leveraging the QSS dense polynomial approximations of the particle trajectories. The sequence diagram in Figure \ref{fig_crossing_algorithm} captures the most relevant interactions of \retQSS' boundary crossing algorithm. 

Suppose we are interested in finding the next boundary crossing of a traveling particle $p$. To accomplish this, the face crossing algorithm first queries the Particle Tracker to retrieve $p$'s current volume, $v$. Particle $p$ knows its \textsl{trajectory}, modeled as a C++ object that stores the polynomial coefficients (supplied by QSS Solver's engine) that approximate the actual trajectory. If volume $v$ has a small number of faces (i.e., less than a configurable threshold $F$, usually set to 50), the algorithm will successively test each of them for possible intersections, invoking for this purpose the \textsl{Intersection Locator} object. Given a face $f$, the Intersection Locator solves a polynomial equation involving the plane that contains $f$ and the approximation of $p$'s trajectory. When such equation is satisfied, it is tested whether the candidate point $P$ yielded by its solution belongs to $f$. In this case, point $P$ is an actual intersection point through exit face $f$. In any other situation (i.e., the equation cannot be solved or no candidate point belongs to $f$), the face crossing algorithm dismisses this face. After discovering an exit face, the associated exit time $t_f$ is used to update the minimum exit time $t_{\textrm{min}}$ found so far. Finally, after all faces are tested, the crossed face data associated with $t_{\textrm{min}}$ is returned.

\begin{figure}[h]
    \centering
    \includegraphics[scale=0.15]{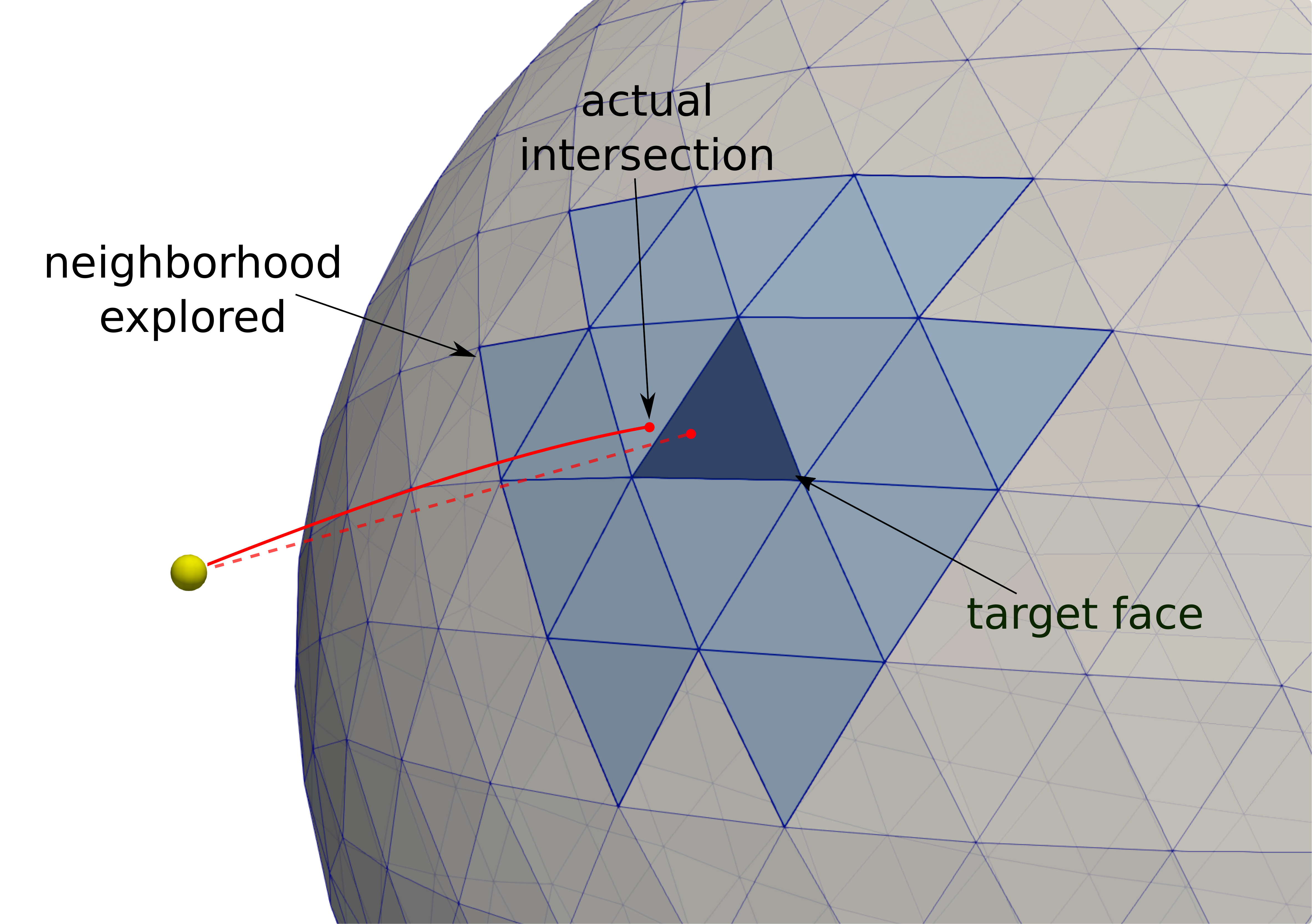}\par 
    \caption{Sketch of boundary crossing detection in volumes with many faces}
    \label{fig_intersection}
\end{figure}

When the number of faces in $v$ exceeds $F$ (not shown in Figure \ref{fig_crossing_algorithm}), instead of testing each face, the algorithm first finds a \textsl{target face} $f'$ crossed by a ray sourced at $p$ with direction given by $p$'s velocity vector. This is efficiently accomplished using $v$'s underlying AABB tree (as explained in Section \ref{sec:geometry}). Then, a small neighborhood of face $f'$ is explored, testing each neighboring face $f$ as explained above. A face $f$ is considered neighbor of $f'$ iff there is a sequence of $1 \leq k \leq d$ faces $f' = f_1,\dots,f_k = f$ such that $f_{i+1}$ is  adjacent to $f_i$ (i.e., they both share one edge), $1 \leq i < k$. Here, $d$ is the \textsl{face neighborhood depth}, fixed at compilation time. This algorithm is particularly important for enabling particles to exit and eventually re-enter the geometry, as the number of boundary faces is usually large. Figure \ref{fig_intersection} illustrates this scenario. 

Boundary crossing detection is a key concept in HEP particle simulations. In Section \ref{sec:hep} we will test our algorithms in the context of a simple HEP setup, comparing their performance against Geant4, the de-facto simulation toolkit for such experiments.

\subsection{Model API}
\label{sec:model_api}

\begin{figure}[h]
    \centering
    \includegraphics[width=0.8\textwidth]{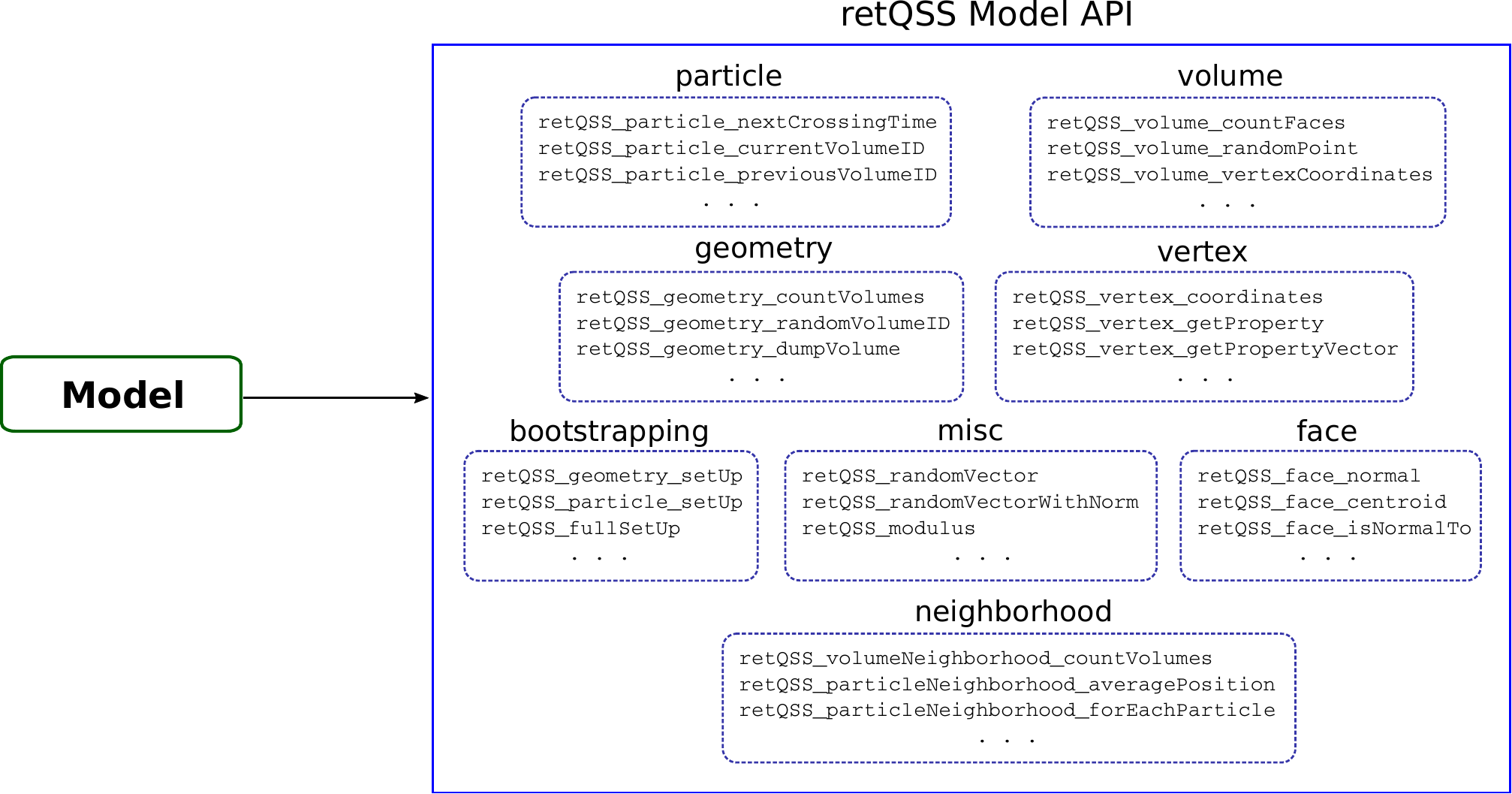}\par 
    \caption{Model API queries and categories}
    \label{fig_model_api}
\end{figure}

The Model API is a public interface that serves as the entry point to \retQSS. This interface exposes several \textsl{queries} implemented by C functions that are organized in eight different categories, as shown in Figure \ref{fig_model_api}.

\textsl{Bootstrapping} queries initialize the required internal data structures and should be called first. Models can initialize geometry and particles in a decoupled fashion. This is particularly convenient to define initial particle conditions using the geometry, as will be illustrated by the bouncing balls model in Section \ref{sec:bouncing_balls_revisited}. For example, particles can be randomly placed in the geometry computing first a random volume ID and then finding a random point inside this volume. The former action is an example of a \textsl{geometry} query. Other geometry queries can be used for debugging or visualization purposes (e.g. exporting volumes to VTK files or dumping geometry statistics). 

Each \retQSS first-class object has its own category, typically providing property access mechanisms and other object related behavior (e.g. retrieving the spatial coordinates of a vertex or computing the outward normal vector of a face). 

One important particle query is \texttt{nextCrossingTime}, which calculates the time of the first upcoming volume boundary crossing of a given particle and updates tracking information when such volume crossings are detected. This particle query is usually employed to react to boundary crossings and trigger a discrete event to handle this action. The zero-crossing function controlling the occurrences of such events compares the simulation time against the next crossing time computed by this query. This way, \retQSS handles the \textsl{state events} originated by the intersection between particle trajectories and volume boundaries as \textsl{time events} formulated in terms of the time instants at which these intersections occur. The combination of the expressive power of \uModelica~with the efficient boundary crossing detection algorithm covered in Section \ref{sec:crossing_algorithm} not only allows for such conversions to happen transparently but also for improved performance, as the localization of state events tends to be computationally demanding \cite{Cellier2006ContinuousSimulation}.

The \textsl{neighborhood} category groups queries related to both volume and particle neighborhoods. \retQSS offers queries for some standard particle neighborhood operations such as computing the average position or velocity of the neighbors. New operations can be defined using neighborhood iteration queries. These traverse the neighborhood and apply any custom function to every particle found, thus extending the high-level modeling language (to this end, advanced users must implement a C/C++ function that captures the desired interactions between neighboring particles and supply its name as a query parameter). For every particle in the neighborhood, \retQSS assembles a \textsl{neighbor bundle} that encapsulates the source particle, its neighbor and their related data (e.g., shortest vector joining them) and invokes the new function. The output value is accumulated and returned to the model once the neighborhood is fully traversed.

The Model API is exported in a Modelica package where each \retQSS query is declared as an external function. Models import this package before interacting with the \retQSS engine.

\section{Bouncing balls example revisited}
\label{sec:bouncing_balls_revisited}

\begin{figure}
    \centering
    \includegraphics[width=0.7\textwidth]{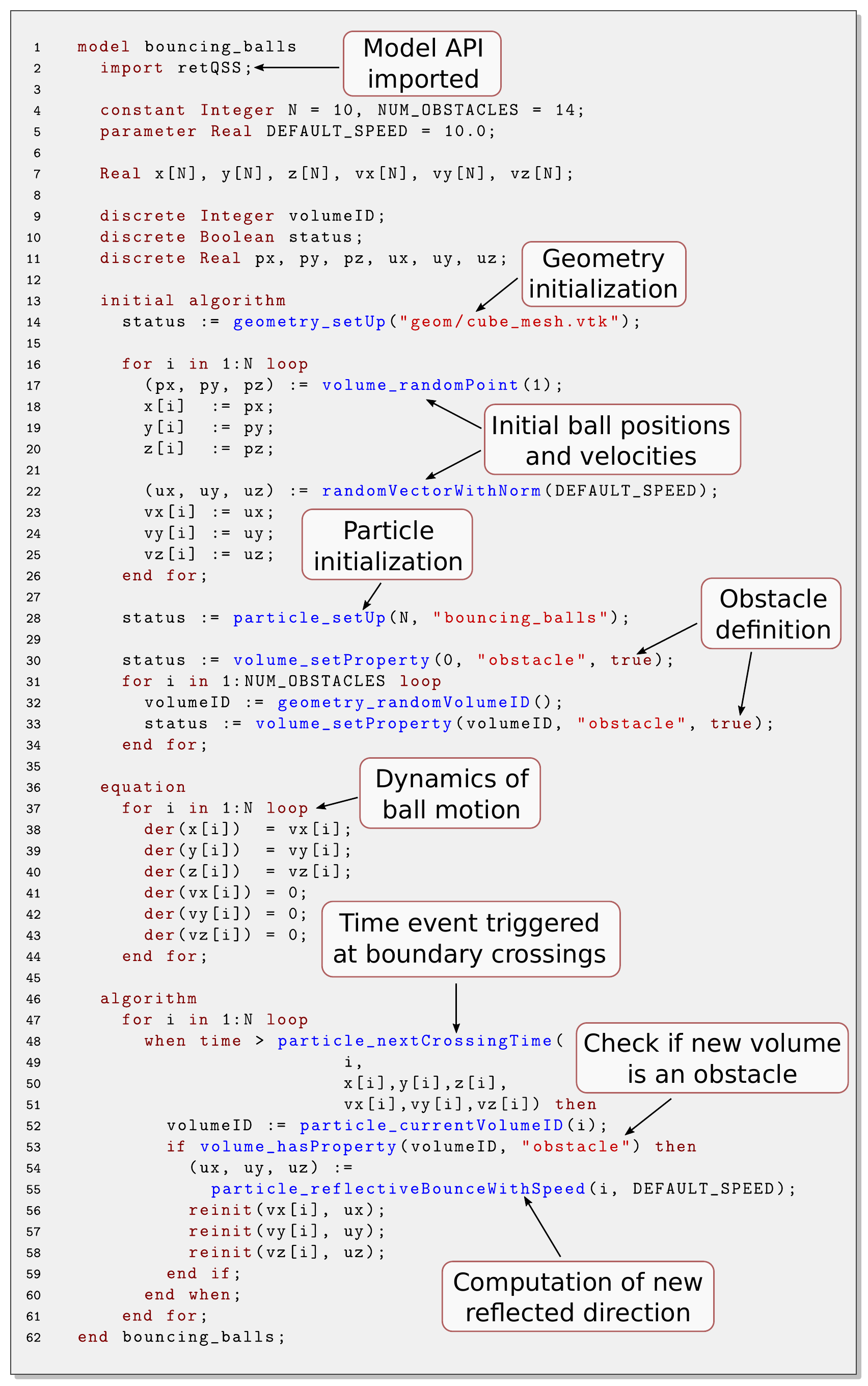}\par 
    \caption{Bouncing balls model}
    \label{fig_bouncing_balls_model}
\end{figure}

Having covered the essential technical concepts and inner workings of \retQSS, we are now ready to discuss how to bring everything together to create a \retQSS model. To this end, we will revisit the bouncing balls example introduced in Section \ref{sec:motivation} and explain in detail one possible implementation, shown in Figure \ref{fig_bouncing_balls_model}.

One of the first actions every model should undertake is to import the \retQSS package that contains the Modelica interface to the model API (line 2). After this point, the model is ready to execute \retQSS queries. This happens first in line 14, where the model issues a bootstrapping query to initialize the underlying geometry and related data structures (supplied as a VTK file). The \texttt{for} loop in lines 16-26 sets the initial conditions (i.e., position and velocity) for each of the \texttt{N} balls in the model (as indicated by line 4, there are 10 such balls). Balls are initially positioned in random locations inside volume 1 (lines 17-20). Their initial directions are also chosen at random, setting for this purpose random velocity vectors with a fixed norm --a \textsl{default speed} that is constant throughout the simulation (lines 22-25). Once these initial conditions are defined, the model calls another bootstrapping query in line 28 so that \retQSS can create particle objects and locate them in the geometry using this information. Before starting the simulation, some volumes are declared as \textit{obstacles}. To accomplish this, we define the \texttt{obstacle} volume property in \texttt{NUM\_OBSTACLES} random volumes (lines 30-34). Note that volume 0 is also marked obstacle (line 30) so that balls bounce back when they hit a boundary face.

The uniform linear equations of motion for the balls are described by the ODE system in lines 37-44. Balls will only modify their directions when bouncing against an obstacle. The system is fully integrated by the standard QSS Solver's simulation engine. Interactions with \retQSS occur according what is defined in the discrete portion of the model in the \texttt{algorithm} section (starting in line 46). A discrete event is defined for each ball (lines 48-60) to handle volume crossings. QSS Solver keeps track of the corresponding zero-crossing functions (line 48) to decide when the events should be triggered. At the heart of this zero-crossing function is the boundary crossing detection algorithm presented in Section \ref{sec:crossing_algorithm}. When such a boundary crossing is detected, the event handler is executed (lines 52-59). In case the new volume is an obstacle (i.e., property \texttt{obstacle} is defined), the ball bounces back with a direction given by the reflection of its current velocity vector across the normal vector of the crossed face. The \texttt{reflectiveBounceWithSpeed} particle query (line 55) computes this new direction. Finally, ball direction is updated by restarting the velocity state variables with these values (lines 56-58).

\begin{figure}[h]
    \centering
    \includegraphics[width=0.8\textwidth]{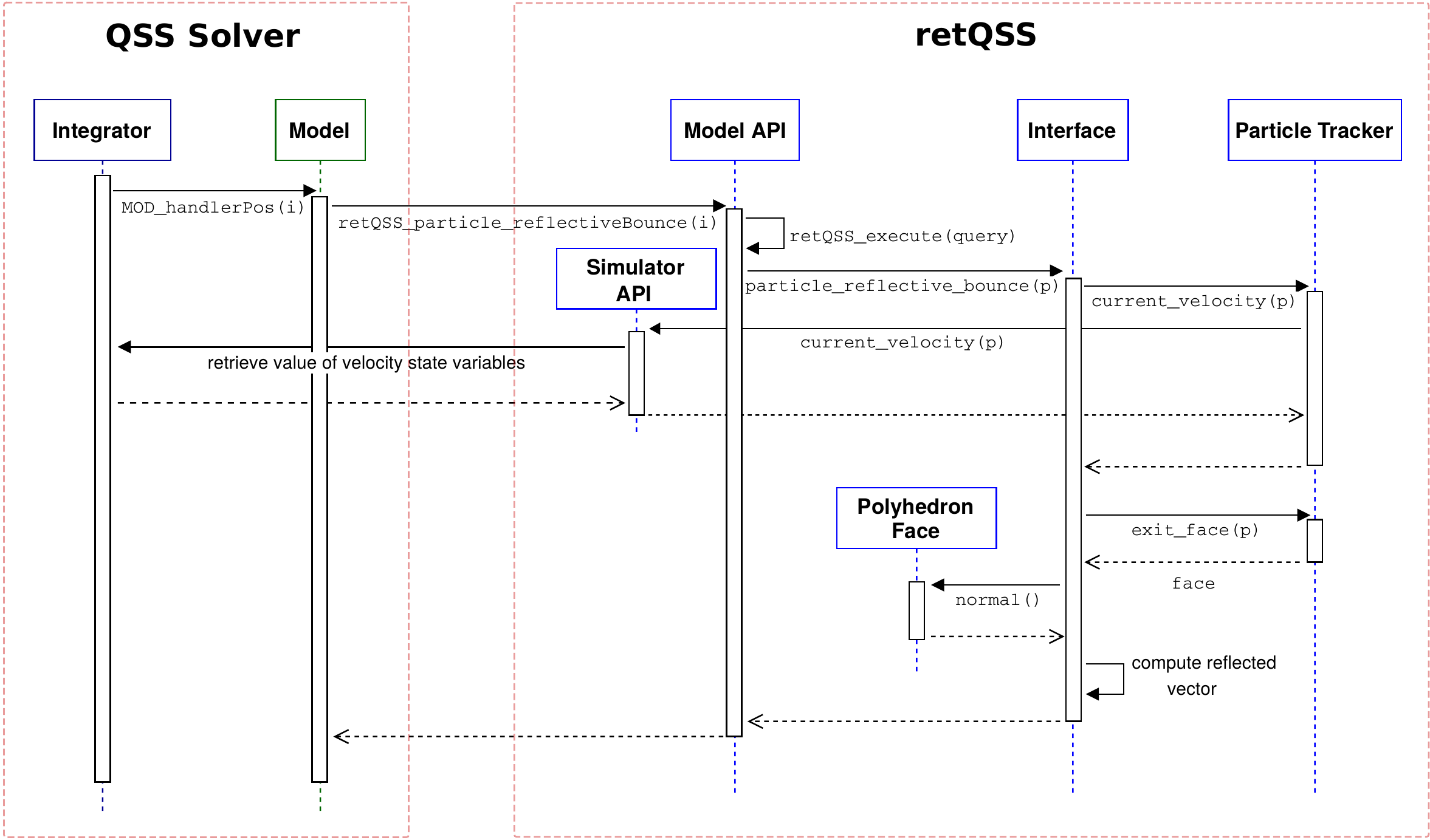}\par 
    \caption{Execution path of the \texttt{reflectiveBounce} particle query}
    \label{fig_reflective_bounce}
\end{figure}

For a better understanding of the execution path of a \retQSS  query, Figure \ref{fig_reflective_bounce} shows how QSS Solver (in particular, the Integrator and Model modules) and \retQSS interact with each other to compute the reflected direction of a particle (i.e., a ball) upon crossing a volume boundary. This action occurs inside an event handler that is executed at some point of the main integration routine of QSS Solver. As discussed in Section \ref{sec:solver}, events are defined and implemented in the Model module. Thus, the Integrator forwards the execution of the event handler to the Model. There, the \texttt{reflectiveBounce} particle query is relayed to the Model API, which then forwards its execution to the internal Interface. The Particle Tracker is queried twice to retrieve the velocity vector $\vec{v}$ of the particle and the face that was just crossed. The former request involves querying back the simulation engine to obtain the values of the velocity state variables. Afterwards, the face calculates its normal vector $\vec{n}$ (in particular, it is an outward-facing unit normal vector), and finally the reflected vector $\vec{r}$ is computed with the reflection formula $\vec{r} = \vec{v} - 2 \vec{v} \cdot \vec{n}$.

\section{Selected case studies and applications}
\label{sec:case_studies}

The purpose of this Section is to develop a comprehensive study of the modeling and simulation capabilities of \retQSS. To this end, we shall address four problems coming from very different application domains. We start by discussing how the features of \retQSS can be leveraged to express versatile models of bird flocking, characterizing birds as an agent system with emergent behavior. We then move to the high-energy physics (HEP) domain, where we will model a charged subatomic particle describing helical trajectories in a mesh of cuboids. In order to assess the performance of our approach, we shall compare it against an equivalent model implemented in Geant4, the reference simulation toolkit used in HEP experiments. Finally, we show how particle-based numerical methods can be efficiently expressed in \retQSS. To this end, we tackle two scenarios: a system of molecules interacting via an exponentially decaying potential and the flow of plasma in a two-dimensional grid. These two classic examples showcase the capabilities of the Molecular Dynamics (MD) and Particle-In-Cell (PIC) methods, respectively. In these cases, we will also carry out a performance study comparing the approaches of \retQSS against those of Aboria (a general-purpose software library for particle-based methods) and Octave (a general purpose toolkit for numerical computing).

A complete description of the hardware and software platform used throughout the experimentation is provided in \ref{sec:platform}. The underlying experimental data and \retQSS models can be found in \cite{ExperimentData}. Animated visualizations of some of these case studies are provided as supplementary material.

\subsection{Bird flocking}
\label{sec:boids}

The \textsl{boids} model \cite{Reynolds1987FlocksModel} is perhaps the most popular flocking model. With applications in many fields such as computer graphics, unmanned vehicle guidance and virtual reality, this model is a classical example of an agent system with complex behavior that emerges from the interactions between individual agents. In other words, the flock behavior is a consequence of the motion and the interactions of the so-called boids --\textsl{bird-oid} objects.

The model is based upon three local rules that capture the steering behavior of a given boid with respect to the positions and velocities of its neighboring flockmates:

\begin{numcases}{}
    \textrm{\textbf{(Separation)}} \quad S(i) = -\sum_{j \, \in \, \Flock{i}}{(\vecFromTo{i}{j})}  \label{eq:boids_sep} \\
    \textrm{\textbf{(Cohesion)}} \quad\,\,\,\, C(i) = \left( \frac{1}{f_i} \sum_{j \, \in \, \Flock{i}}{\PartPos{j}} \right) - \PartPos{i}  \label{eq:boids_coh} \\
    \textrm{\textbf{(Alignment)}} \quad A(i) = \frac{1}{f_i} \sum_{j \, \in \, \Flock{i}}{\PartVel{j}} \label{eq:boids_align}
\end{numcases}

Here, $\PartPos{i}$ and $\PartVel{i}$ represent the position and velocity of boid $i$, respectively, $\Flock{i}$ is the set of its neighboring flockmates and $f_i$ is the size of such set.

The \textsl{separation} rule (Equation~\eqref{eq:boids_sep}) makes a boid steer away from its local flockmates, thus avoiding collisions. The \textsl{cohesion} rule (Equation~\eqref{eq:boids_coh}) is the steering force through which a boid moves towards the center of its visible flock, thus allowing the formation of clusters. Finally, the \textsl{alignment} rule (Equation~\eqref{eq:boids_align}) makes a boid match its heading with the average heading of its neighbors, thus allowing for an emerging pattern of motion. 

When the heading $\PartDir{i}$ of boid $i$ is updated, its new value is given by a weighted sum of the headings produced by each of these rules,
$$
\PartDir{i} := w_S \cdot S(i) + w_C \cdot C(i) + w_A \cdot A(i)
$$
In typical boids implementations, weights $w_S, w_C, w_A$ are model parameters. Also, it is generally assumed that boids follow uniform linear motions, periodically updating their headings as described above. The set of neighboring flockmates of boid $i$, $\Flock{i}$, is usually conformed by every other boid $j$ such that the Euclidean distance between them is no greater than a parameter $r$:
$$
\Flock{i} = \left\{ j \neq i : |\vecFromTo{i}{j}| \leq r \right\}
$$
In what follows we will explore two different strategies to model this system in \retQSS. Our proposed modeling strategies share a common structure, differing in how the headings are updated. This structure can be summarized as follows:
\begin{itemize}
    \item Exploiting the capabilities of the Model API, boids are transparently placed in random initial positions throughout the geometry. 

    \item Local flocks are represented by particle neighborhoods. In turn, these are supported by radial volume neighborhoods with radius $r$.
    
    \item Boid trajectories are modeled with a straightforward ODE system like the one explored in Section \ref{sec:bouncing_balls_revisited}.
    
    \item The boundary faces in the geometry are treated as rigid walls. Boids bounce and remain inside the geometry after hitting one of these faces.
    
    \item Heading updates are expressed as a discrete time event that is triggered after a regular interval with a configurable duration. The handler of this event successively updates every boid in the system according to the underlying update strategies discussed below.
\end{itemize}

\begin{figure}[h]
    \begin{center}
    \begin{tabular}{cc}
      \subfloat[Torus\label{fig_boids_torus}]
        {
            \includegraphics[scale=0.085,valign=c]{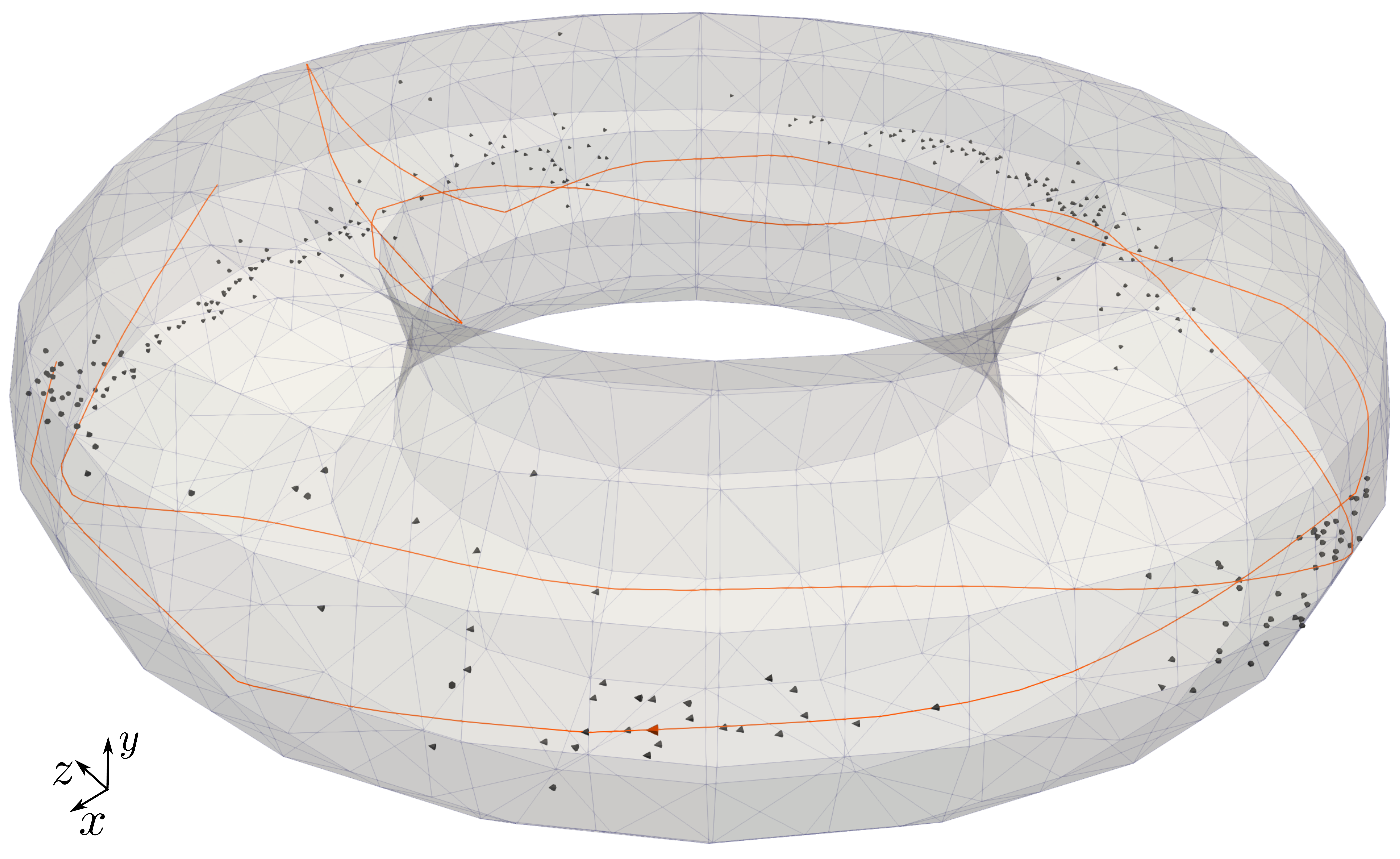}
        }   &
      \subfloat[Sphere\label{fig_boids_sphere}]
        {
            \includegraphics[scale=0.11,valign=c]{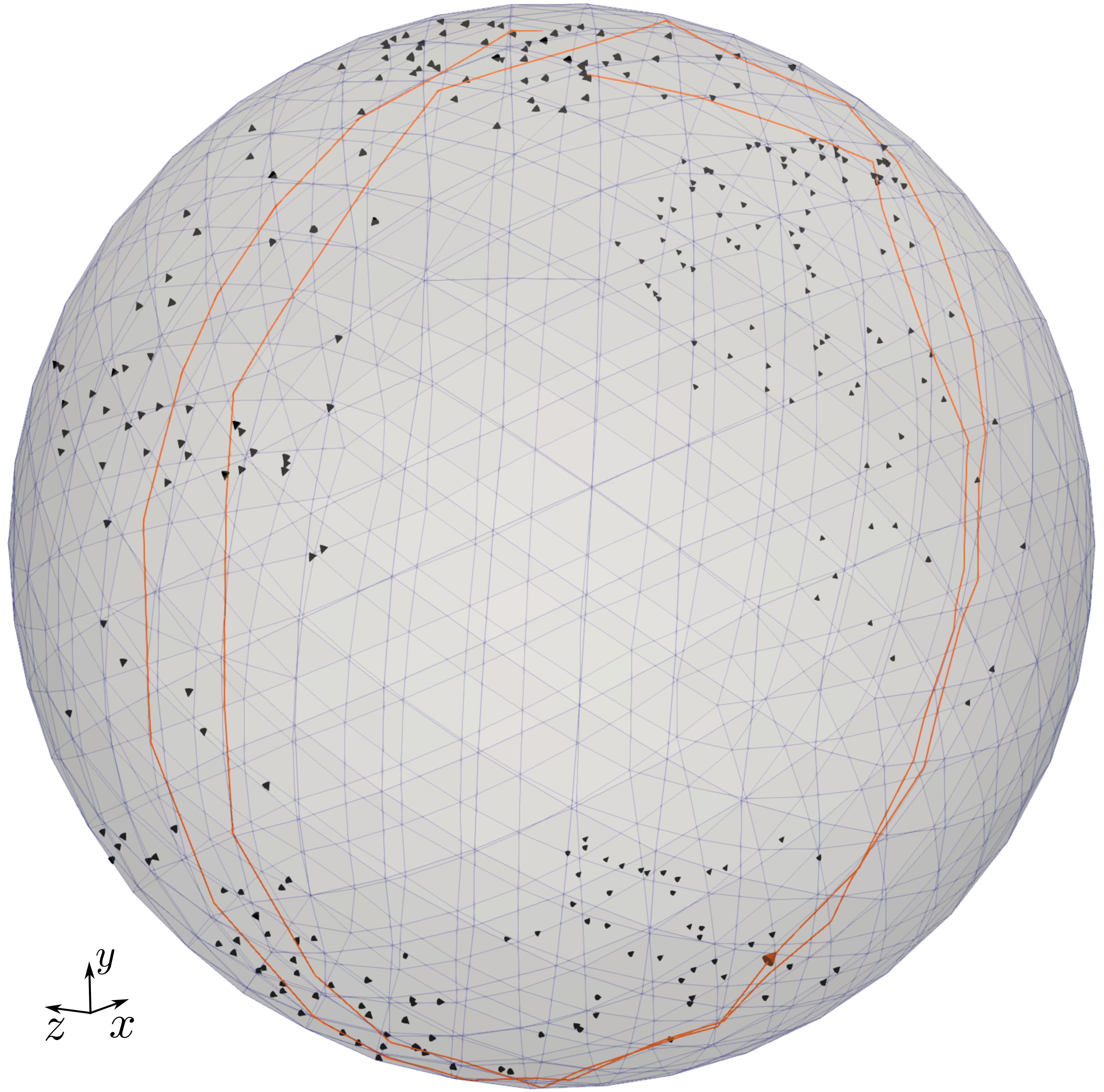}
        }    
    \caption{Bird flocking in two alternative geometries (sample trajectories highlighted)}
    \label{fig_boids}
    \end{tabular}
    \end{center}
\end{figure}

We stress that the mechanics to implement these strategies are geometry-agnostic. This is illustrated by Figure \ref{fig_boids}, which shows two boids simulations in the context of very different geometries (tessellated torus and sphere). Note that, in case the extents and dimensions of such geometries are not comparable, the user might also need to update the radius parameter after updating the input geometry. 

\subsubsection{First approach: using external functions}
\label{sec:boids_external}

\begin{figure}[h]
    \centering
    \includegraphics[width=0.6\textwidth]{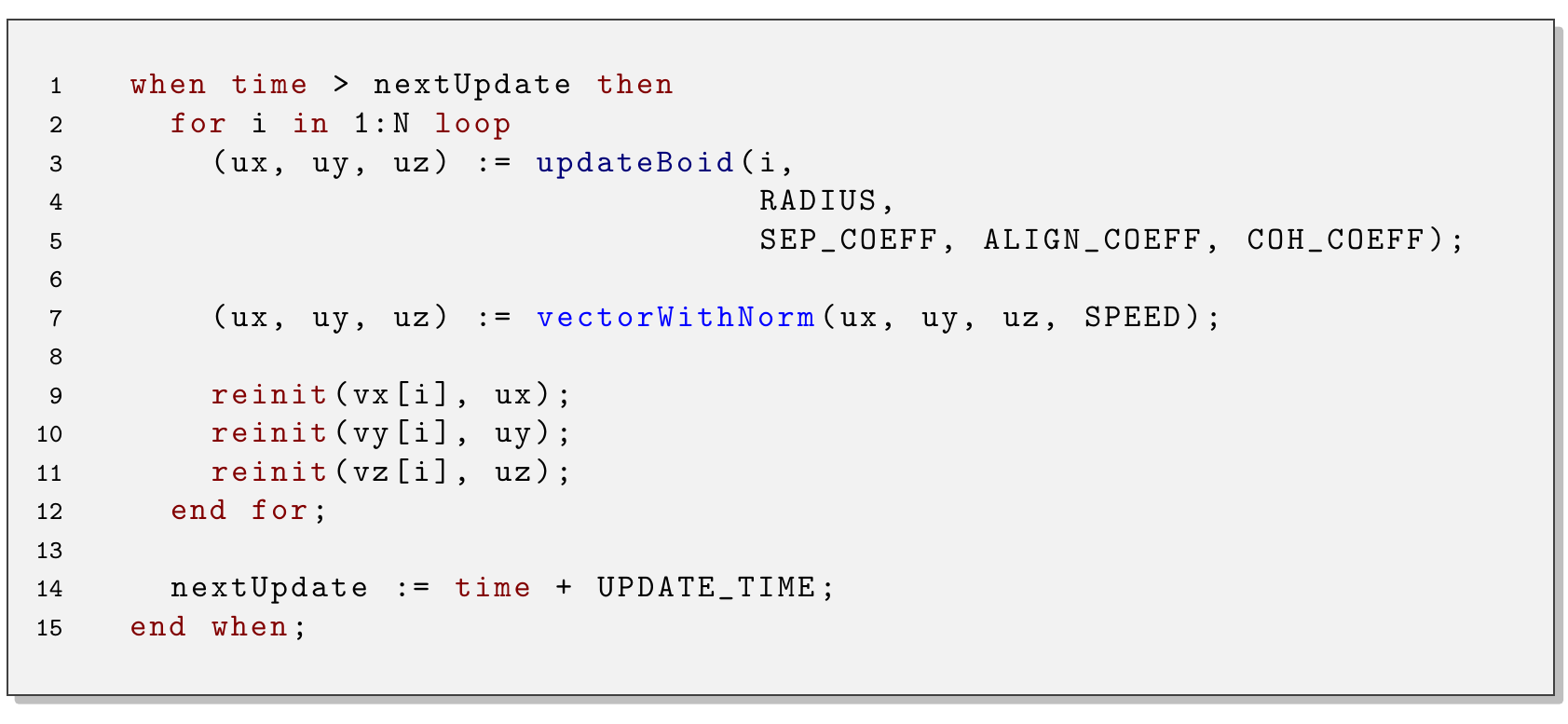}\par 
    \caption{Boids headings updated using an external C++ function}
    \label{fig:boids_ext}
\end{figure}

Our first approach uses a custom external C++ function to evaluate the rules and compute the new heading vectors. Figure \ref{fig:boids_ext} shows an excerpt of this model with the definition of the time event that triggers the computation of the new headings.

The external function \texttt{updateBoid} (line 3) is exposed to \uModelica~in a user-defined package that has to be imported in the main model file. The public C interface exposed by the model API of \retQSS  is an essential component behind this approach, as rules are computed by traversing particle neighborhoods using C++ code. Given a boid $i$, the function starts by initializing three 3D vectors (exposed by \retQSS via the CGAL library) to represent the headings of the behavioral rules of the model. The source volume of boid $i$, along with each of its corresponding neighboring volumes, are processed one at a time by a helper function. This function iterates the particles inside the given volume and updates the three heading vectors for every visited particle within the target distance $r$ (parameter \texttt{RADIUS}). The new heading is finally computed by evaluating the weighted sum using the given coefficients (parameters \texttt{SEP\_COEFF}, \texttt{ALIGN\_COEFF} and \texttt{COH\_COEFF}).

The main advantage of this strategy stems from the ability to seize the expressive power and efficiency of a programming language such as C++ to implement small portions of the modeling logic. In some circumstances, achieving the same outcome through a high-level modeling language can be challenging. Yet, this comes at a cost. On the one hand, the modeler may need some programming skills to fully leverage the benefits of this approach. Also, it may not be desirable to break down the model description across different languages. 

\subsubsection{Second approach: using particle neighborhoods}
\label{sec:boids_neighborhoods}

\begin{figure}[h]
    \centering
    \includegraphics[width=0.6\textwidth]{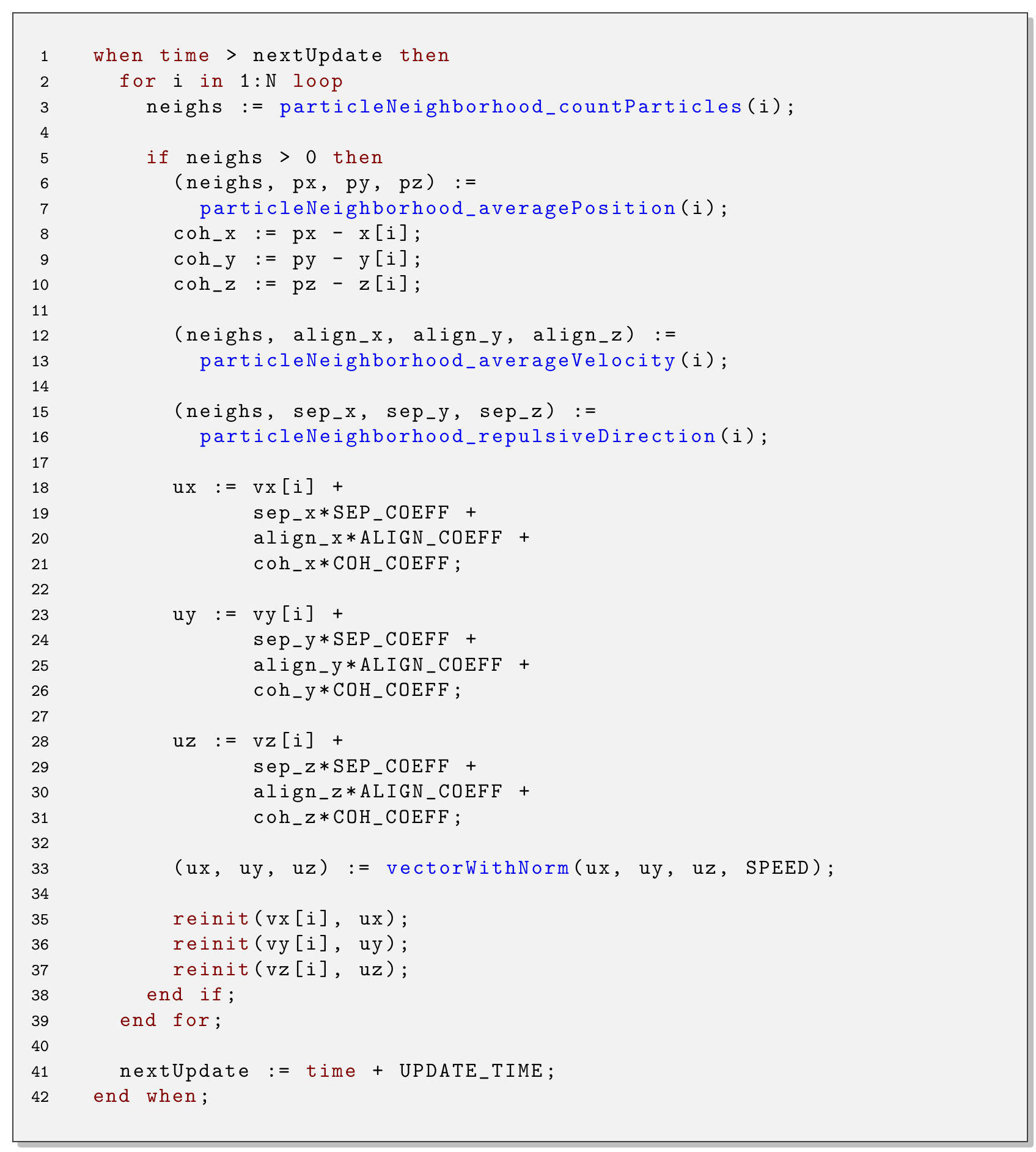}\par 
    \caption{Boids headings updated using particle neighborhood queries}
    \label{fig:boids_funcs}
\end{figure}

The particle neighborhood API provides some default convenience queries well suited for this example. The center of mass of the local flockmates of a boid (a central concept in the cohesion rule) is precisely the result of the \texttt{averagePosition} particle neighborhood query. Similarly, the alignment rule involves the computation of the average heading of neighboring boids, which can be inferred from the resulting vector of the \texttt{averageVelocity} query. Finally, the \texttt{repulsiveDirection} query yields the average of the directions opposite to each of the neighbors, which coincides with the heading specified by the separation rule. We can thus capitalize these queries to succinctly express the heading update logic in \uModelica. This is shown in Figure \ref{fig:boids_funcs}.

Another hybrid implementation that mixes both approaches is also possible. As discussed in Section \ref{sec:model_api}, the particle neighborhood API includes a query that automatically traverses a particle neighborhood and applies a custom C/C++ function to every particle found. Should the modeler choose to follow this path, he or she would need to implement an external function to capture the heading update logic for the three rules. This function would receive a neighbor bundle object connecting the source boid $p$ with a neighbor $q$. In other words, this strategy would spare the modeler to code the neighborhood traversal logic explored before.

\subsubsection{Possible extensions}
\label{sec:boids_extensions}

The boids model can be easily extended in several ways leveraging the features showcased by \retQSS. For example, we may define new rules to control the steering behavior of the boids based on external forces such as wind. Using vectorial volume properties, it becomes straightforward to define arbitrary directions and intensities of the wind. It is also possible to provide more complex characterizations. Much like the continuous dynamics of particle trajectories, we might capture wind behavior in a set of mathematical equations and later use them to periodically update the volume properties.

Another interesting extension is the introduction of obstacles in order to investigate obstacle avoidance mechanisms. By following a modeling approach similar to the bouncing balls example (Section \ref{sec:bouncing_balls}), an obstacle can be represented by a volume featuring an obstacle property. Then, we can consider a simple avoidance strategy consisting in inverting the heading of a boid when it faces an imminent obstacle. One suitable detection technique is testing whether the volume corresponding to the next scheduled boundary crossing is an obstacle. Indeed, the particle query \texttt{nextVolumeID} returns the ID of such volume. 

\subsection{High-energy physics}
\label{sec:hep}

Particle simulation in the HEP domain typically involve tracking subatomic particles affected by physics processes as they interact with matter traveling across complex detector geometries. Such geometries are modeled as a composition of three-dimensional volumes of potentially different shapes and materials. The accuracy and performance of the underlying particle tracking algorithms are of great interest since they can have a considerable impact in the requirements for computing resources and their associated cost. In particular, these algorithms need to deal with frequent discontinuities caused by the recurrent crossing of geometrical boundaries by a traveling particle, from one volume to the next.

The purpose of this Section is twofold. Although we are mainly interested in demonstrating the flexibility of \retQSS by tackling a typical HEP problem, we also intend to show the efficiency of our particle tracking methods. We will focus on a simple model and establish a performance comparison against Geant4, the most widely adopted simulation toolkit in modern HEP experiments \cite{Allison2016Geant4}.

The model consists of a charged particle under the action of a static magnetic field in a mesh of cuboids. In spite of its apparent simplicity, this model exhibits interesting features. First, it has a closed-form analytic solution which enables an accurate error analysis. In addition, the number of boundary crossings can be easily controlled, as it is directly linked to the edge size of the cuboids. Different variants of this model were used in the past to characterize the performance of QSS-based numerical integration strategies specially developed for Geant4 \cite{Santi2019APhysics,Santi2020HEP}.

\subsubsection{Model description}
\label{sec:hep_model}

The model represents a single positron under a uniform, static magnetic field along the $\hat{z}$ plane. The equations that govern the motion of a charged particle in a magnetic field (Lorentz equations) are presented in Equation~\eqref{eq:lorentz}. There, $q$ and $m$ stand for the charge and mass of the particle, respectively; $c$ is the speed of light; $\gamma$ is the Lorentz factor and $\vec{B} = (B_x, B_y, B_z)$ is the magnetic field.

\begin{equation}
\systeme{\dot{x} = v_x \quad\quad \dot{v_x} = \frac{q \, c^2}{m \, \gamma} \cdot (v_y \, B_z - v_z \, B_y),
          \dot{y} = v_y \quad\quad \dot{v_y} = \frac{q \, c^2}{m \, \gamma} \cdot (v_z \, B_x - v_x \, B_z),
          \dot{z} = v_z \quad\quad \dot{v_z} = \frac{q \, c^2}{m \, \gamma} \cdot (v_x \, B_y - v_y \, B_x)
\label{eq:lorentz}}    
\end{equation}\\

In our case we set $B_x = B_y = 0$, i.e., $\vec{B} = (0, 0, B)= B \hat{z}$. The particle has an initial velocity $\vec{v} = (w \cdot v\hat{x}, 0, \sqrt{1 - w^2} \cdot v\hat{z})$ and a constant speed $v = 0.999 \, c$. $B$ and the coefficient $w$ are two model parameters that control the intensity of the magnetic field and the initial speed in $\hat{x}$, respectively. 

\begin{figure}[H]
    \centering
    \includegraphics[scale=0.2]{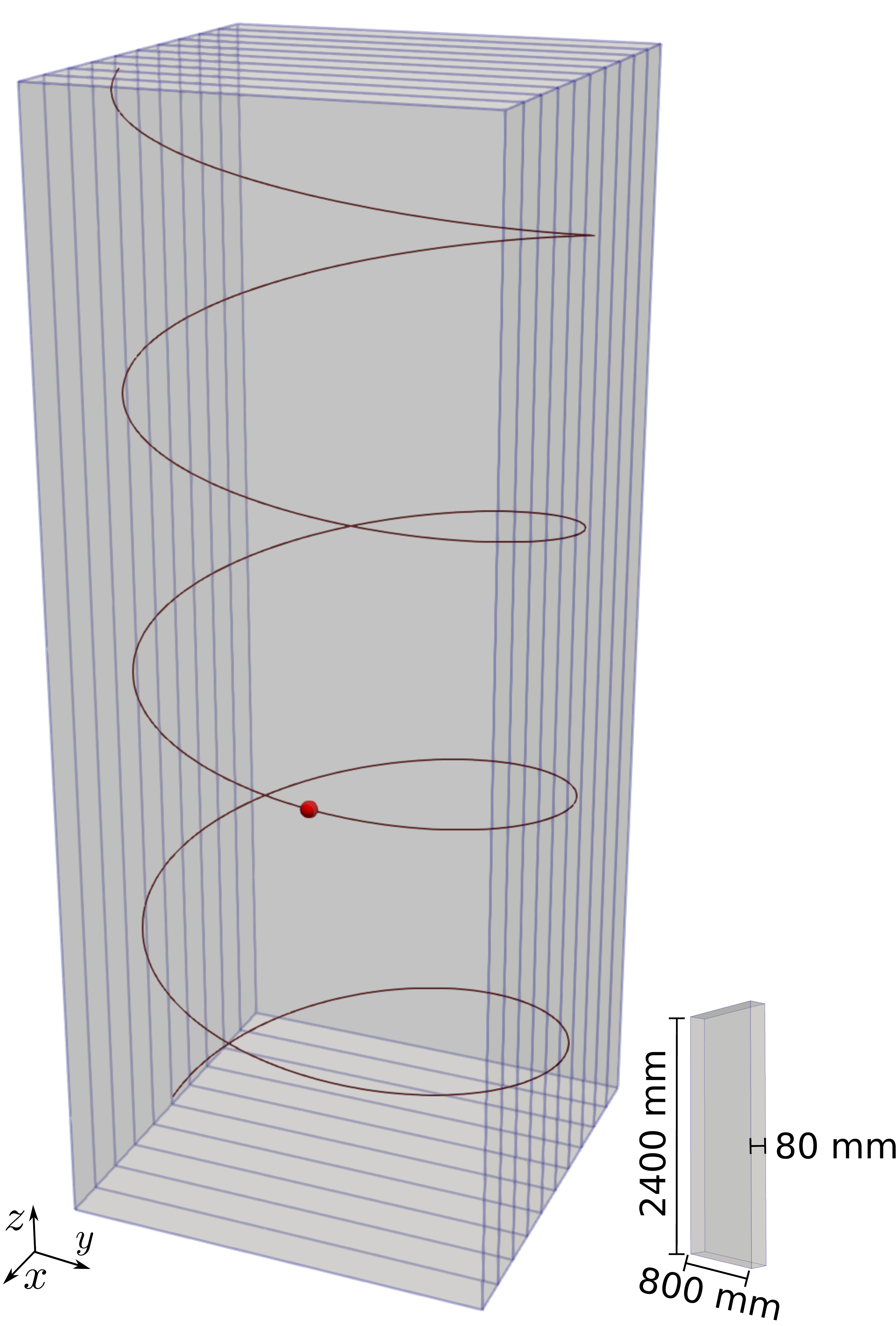}\par 
    \caption{A positron describing a helical trajectory in a mesh of cuboids}
    \label{fig_helix}
\end{figure}

This instance of the ODE system in Equation~\eqref{eq:lorentz} admits an analytic solution. The particle follows a helical trajectory with a linear increase in $\hat{z}$ with respect to time. By choosing $B = 0.0937$ Tesla and $w = 0.98$, the diameter of the trajectory (in $\hat{x}$ and $\hat{y}$) turns out to be about 800 mm. In this situation, we have that after four full revolutions, the particle reaches a maximum height (in $\hat{z}$) of about 3x the diameter. Thus, we can place the particle inside a geometry with these dimensions, as illustrated by Figure \ref{fig_helix}. The number of boundary crossings can be controlled by changing the $\hat{x}$ dimension of the cuboids.

\subsubsection{Implementation}
\label{sec:hep_implementation}

Our \retQSS model provides a high-level Modelica description of the Lorentz equations of motion and uses a discrete time event to detect volume crossings (the related particle query \texttt{nextCrossingTime} allows for a straightforward implementation of this). Each intersection point is stored for postprocessing purposes.

The Geant4 model is based on a standalone C++ application. It provides classes to describe different aspects of the model (e.g. the definition of the magnetic field or the construction of the detector geometry). Numerical integration in Geant4 is performed by its default stepper, a custom implementation of the fifth/fourth-order accurate Dormand-Prince adaptive method (DOPRI745). In \retQSS, we chose QSS2 as the integration method for this case study.

\subsubsection{Validation}
\label{sec:hep_validation}

We validated our model by comparing against Geant4 the absolute error $E$, defined as the \textit{maximum Euclidean distance} between the simulated and the theoretical position of the particle at each timestamp. Using a default relative accuracy $\epsilon = 5.5 \times 10^{-5}$ in Geant4, $E$ represents a $6 \times 10^{-5} \%$ of the total track length of the particle. Setting $\dQRel = 5.5 \times 10^{-5}$ in \retQSS yields an error about 30x higher (with a sufficiently small absolute tolerance, e.g. $\dQMin = 10^{-7}$). Thus, in order to increase the accuracy by a factor of 30, we should reduce the quantum by this same factor, as dictated by one of the QSS properties outlined in Section \ref{sec:qss_properties}. Indeed, with $\dQRel = 1.83 \times 10^{-6}$ we obtain a value of $E$ equivalent to that of Geant4. We carried out the experimentation selecting $\dQRel = 1.6 \times 10^{-6}$ and $\dQMin = 1.6 \times 10^{-7}$. With these values, $E$ is approximately $13\%$ lower in \retQSS.

\subsubsection{Experiment methodology}
\label{sec:hep_methodology}

We experimented with geometries featuring different volume crossing intensities. To this end, we chose 16 equidistant values in the range $[0.5, 2]$ mm for the $\hat{x}$ dimension of the cuboids, $l_x$. Thus, the number of subdivisions of $\hat{x}$ ranges from 1600 (for $l_x = 0.5$ mm) to 400 (for $l_x = 2$ mm). 
For each scenario we extracted two metrics: the \textsl{end-to-end speedup} of \retQSS against Geant4, and the \textsl{boundary crossing overhead} of each simulator. The former is defined as the ratio between the end-to-end simulation times achieved by Geant4 and \retQSS, respectively, excluding initialization processes. The latter is the average CPU time devoted to the computation of boundary crossings. For Geant4, this value is taken as the average CPU time taken by the most time-consuming routine in the boundary crossing detection pipeline. This routine iteratively calculates an intersection point from an initial estimation given by a linear segment that crosses a volume boundary \cite{Santi2019APhysics}. In the case of \retQSS, we start with a baseline scenario $l_0$ that uses a ``hollow'' geometry (i.e., with no subdivisions) to estimate the CPU time required to compute the particle trajectory (we verified, as expected, that the number of internal transitions of the QSS main integration loop remains constant across every scenario $l_x$). Thus, the boundary crossing overhead for scenario $l_x$ is calculated as the difference between the end-to-end simulation times of $l_x$ and $l_0$ divided by the total number of boundary crossings found in $l_x$.  

Each scenario was independently simulated 40 times (20 to compute average end-to-end simulation times and 20 to compute average boundary crossing overheads, as measuring the latter may distort the former). Sample standard deviations remained below 17\%.

\subsubsection{Results and discussion}
\label{sec:hep_results}

\begin{figure}[H]
    \centering
    \includegraphics[scale=0.55]{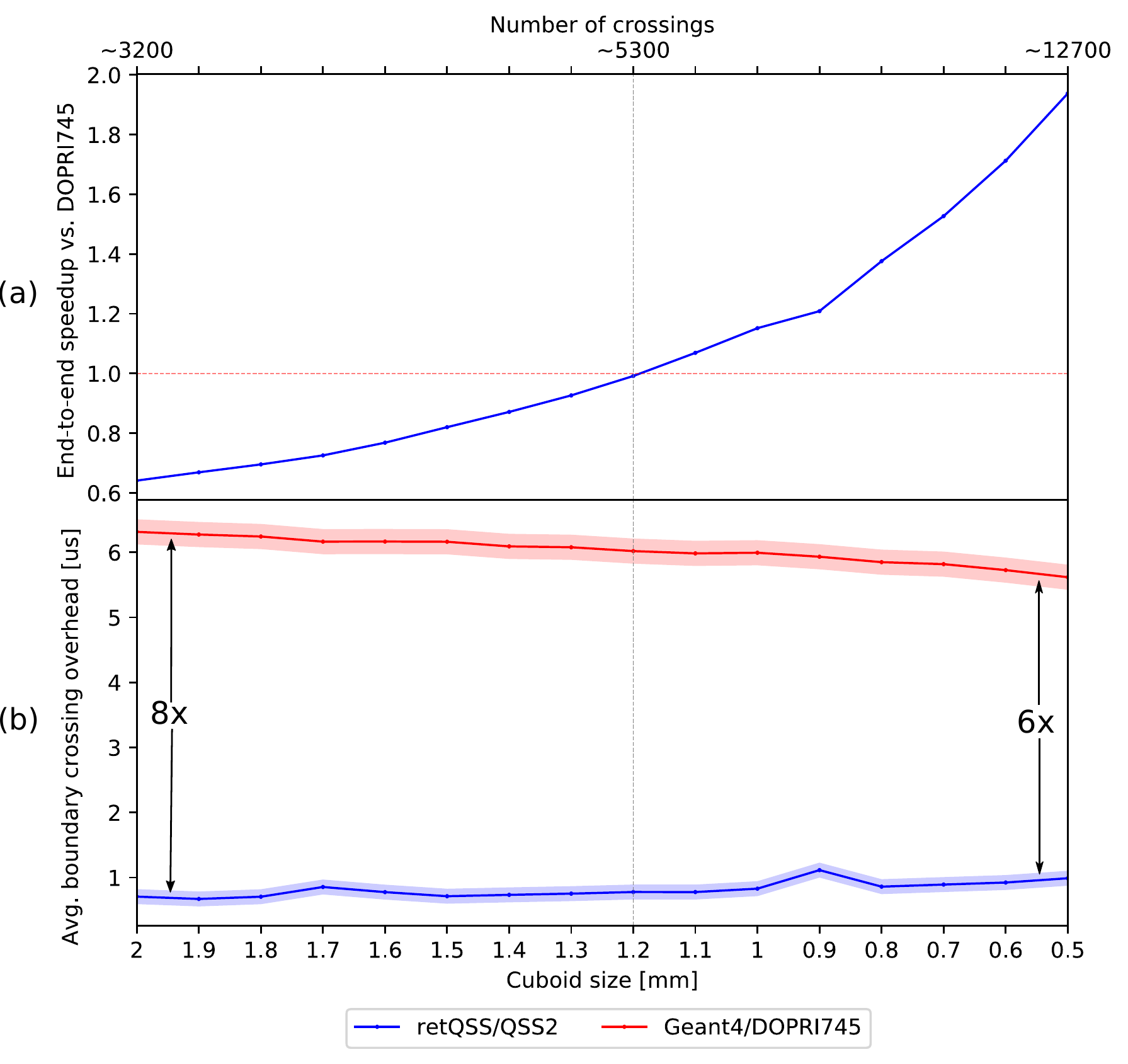}\par 
    \caption{Performance comparison between \retQSS and Geant4}
    \label{fig_hep_perf}
\end{figure}

Figure \ref{fig_hep_perf}a shows that \retQSS consistently improves its efficiency as the cuboid size $l_x$ decreases (which leads to more frequent boundary crossings). Around a value of 1.2 mm (approximately 667 subdivisions of the geometry in the $\hat{x}$ dimension and about 5300 boundary crossings), \retQSS achieves a performance nearly equivalent to that of Geant4. From that point onwards, \retQSS systematically outperforms Geant4, reaching a maximum speedup of about 2x when $l_x = 0.5$ mm.

This can be explained by two facts. First, by observing Figure \ref{fig_hep_perf}b, we see that \retQSS is between 6x and 8x faster than Geant4 to compute boundary crossings. Thus, it is reasonable to observe better performance as the intensity of volume crossings becomes more stringent. However, we should also note that Geant4 demands at least as many computational steps as volumes crossed by the particle during the simulation. When a volume crossing is detected, Geant4 interrupts the stepping routines and ends the step at the volume boundary. Tracking is later resumed in a new step that starts in the neighboring volume. Among other things, advancing a step involves an evaluation of the numerical integration routines, which comes with non-negligible performance penalties. In consequence, we should expect larger simulation times as the number of steps increases.

On the other hand, \retQSS computes the particle trajectory with a fixed cost that does not depend on the number of volume crossings. As we already mentioned, the number of internal transitions of the QSS main integration loop that are due to state changes remains constant across every scenario. Although this cost is very high with respect to a single computational step of Geant4 (about 70x higher, possibly due to the lower order of the underlying numerical solver), the systematic increase in the number of steps mitigates this initial difference.  This remark is not surprising, as it stems from the discrete-event nature of QSS and their consequent ability to deal efficiently with discontinuities in the system (in this case, in the form of boundary crossings).

\subsubsection{Modeling overview}
\label{sec:hep_modeling_overview}

Modeling simple HEP examples such as the one studied in this Section can demand a non-negligible amount of work in traditional environments (due to the usually steep learning curves of toolkits such as Geant4). High-level modeling approaches like the one followed by \retQSS can certainly aid in overcoming this problem. Yet, advanced HEP setups featuring complex geometries (such as a realistic particle detector) may prove difficult to model in \retQSS. 

\subsection{Molecular interactions}
\label{sec:md}

Computer simulations in disciplines such as biochemistry and biophysics are often aimed at studying the movement of atoms and molecules. Molecular Dynamics (MD) is a particle-based numerical method very popular in such disciplines. In a typical MD setup, a system of particles is simulated with discrete time methods, using Newton's laws of motion to update the position and velocity of the particles at each time step. Forces acting on each particle are first computed via a force field that describes the interactions with the other particles in the system. To keep computation under reasonable levels, the method usually implements a cut-off scheme to compute interactions between nearby particles (i.e. those within a certain radius $r$). Efficient mechanisms to compute such force fields are of central importance in software implementations of MD.

In this Section we discuss possible implementations of MD in the context of \retQSS. To accomplish this, we focus on a case study that represents a system of molecules interacting via an exponentially decaying potential. This example is inspired by a similar 2D model proposed as a step-by-step case study to implement MD in Aboria, a general-purpose software library for particle-based methods \cite{Robinson2017Particle-basedAboria}. In order to assess the effectiveness and efficiency of our approach, we also carry out a performance comparison against this tool.

\subsubsection{Model description}
\label{sec:md_model}

\begin{figure}[h]
    \centering
    \includegraphics[scale=0.16]{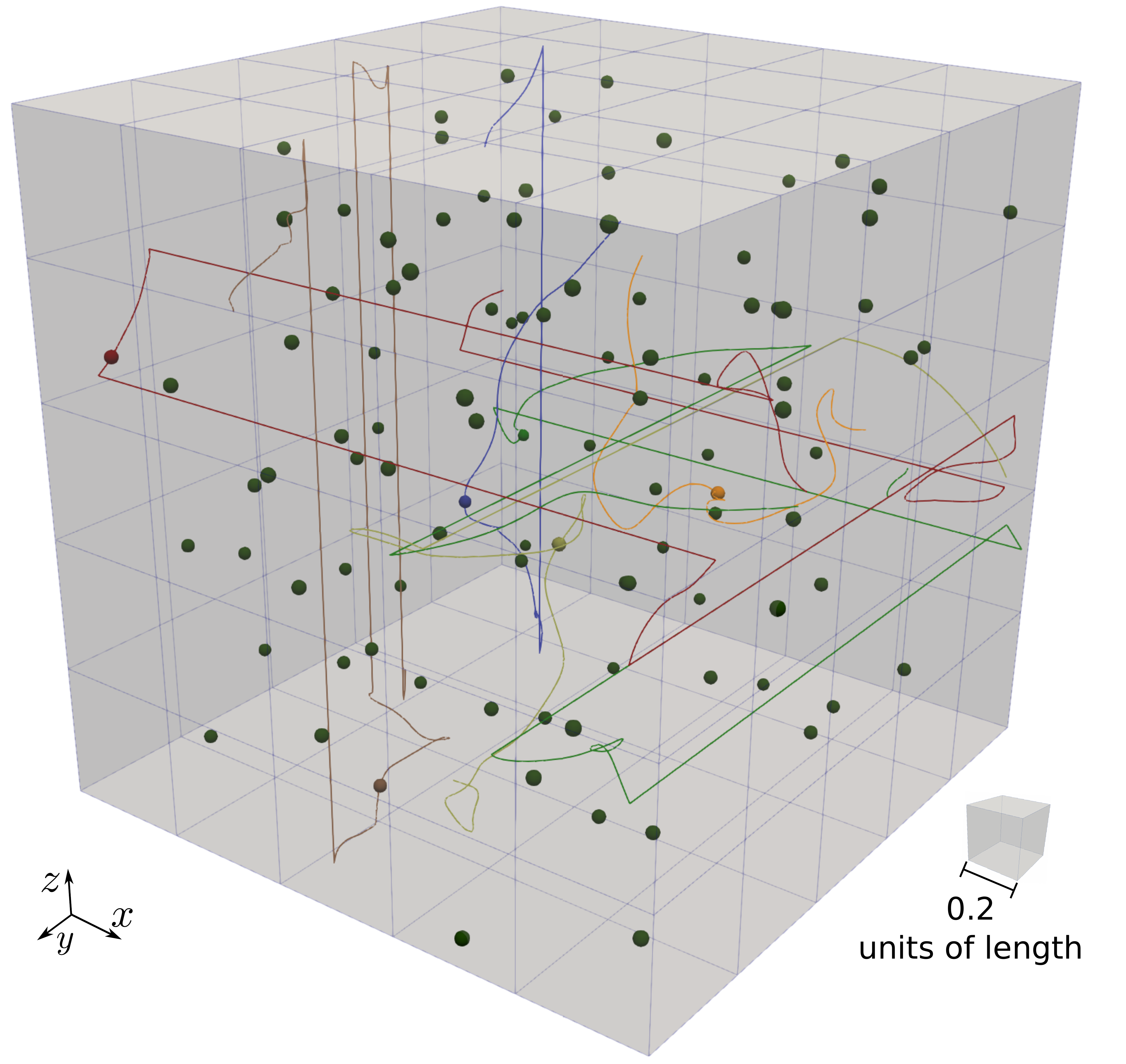}\par 
    \caption{Snapshot of a simulation of molecules interacting via an exponential potential (with highlighted trajectories)}
    \label{fig_md}
\end{figure}

This case study consists of a system of $N = 100$ abstract molecules (with unit mass) within a unit cube with periodic boundary conditions. Initially, every molecule is at rest and randomly located inside the geometry. Intermolecular forces are modeled with an exponential potential with a cut-off value $r_{\textrm{cut}} = 0.2$. The force on molecule $i$ due to molecule $j$, at positions $\textbf{x}_i$ and $\textbf{x}_j$, respectively, is given by
\begin{equation}
\label{eq:force_field}
\textbf{f}_{ij} = 
    \begin{cases}
        -c \cdot \textrm{exp}(-|\textbf{dx}_{ij}|) \, \frac{\textbf{dx}_{ij}}{|\textbf{dx}_{ij}|} & \textrm{if } |\textbf{dx}_{ij}| < r_{\textrm{cut}} \\
        0 & \textrm{otherwise}
    \end{cases}
\end{equation}

where $\textbf{dx}_{ij} = \textbf{x}_j - \textbf{x}_i$ is the shortest vector between $\textbf{x}_i$ and $\textbf{x}_j$ and $c = 10^{-5}$ is a constant. By Newton's second law the acceleration of the $i$-th molecule is given by $\textbf{a}_i = \sum_j{\textbf{f}_{ij}}$. Force fields are evaluated at each of the $T = 2000$ time steps. Figure \ref{fig_md} shows a snapshot of a simulation of this model.

\subsubsection{Implementation}
\label{sec:md_implementation}

In \retQSS, the geometry is meshed with cubes of size $r_\textrm{cut}$. This structure enables the computation of the force field by means of particle neighborhoods supported by periodic radial volume neighborhoods. The chosen radius is $r = r_\textrm{cut} - \epsilon$, where $\epsilon > 0$ is such that the neighbors of a volume are only those volumes around it (i.e., those sharing at least a vertex modulo periodicity). A \textit{time event} triggers the evaluation of the force field for each particle. This is achieved by a particle neighborhood query that takes as input an external C++ function implementing Equation~\eqref{eq:force_field}.

The Aboria model is fully implemented in C++. It employs an ad-hoc semi-implicit Euler integrator embedded in the code. In \retQSS, we chose QSS2 as the underlying integration method.

\subsubsection{Validation}
\label{sec:md_validation}

In order to verify that \retQSS and Aboria produce equivalent solutions, we define a highly accurate baseline scenario in \retQSS  using a very small absolute error tolerance (a fixed quantum of $\deltaQ = 10^{-11}$). We then measure the similarity between both approaches by computing the error \MDErr~taken as the maximum among the Root-Mean Square Deviations (RMSD) of the particle positions at each time step. For random simulations, we found that \MDErr~ typically remains within $0.3\%$ of the extent of the geometry. 

\begin{figure}[H]
    \centering
    \includegraphics[scale=0.45]{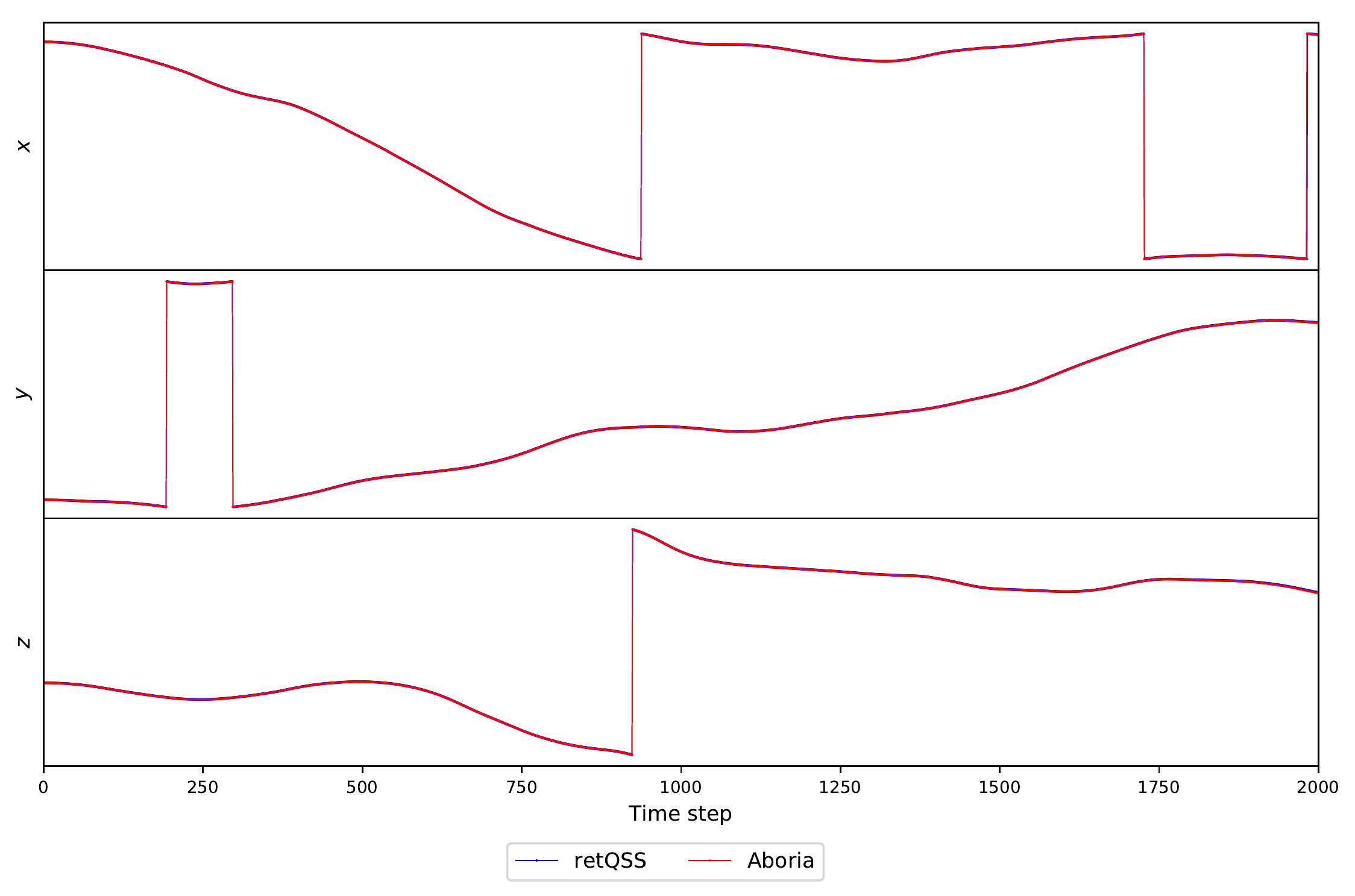}\par 
    \caption{Sample trajectory of a molecule in \retQSS and Aboria simulations (vertical jumps are due to the crossing of periodic boundaries)}
    \label{fig_md_trajectory}
\end{figure}

We also inspected visually the trajectories of the molecules in sample simulations and confirmed that they are indistinguishable to the naked eye, as supported by Figure \ref{fig_md_trajectory}.

\subsubsection{Experiment methodology}
\label{sec:md_methodology}

For a sample simulation in Aboria, we carried out a parameter sweeping for the accuracy in \retQSS and characterized the performance of each scenario.

Performance is characterized by means of a custom \textit{performance index} $\eta$ that combines the simulation time $t_{\textrm{sim}}$ and the error as given by the similarity index \MDErr~ introduced above (Equation \eqref{eq:performance_eta}). This metric enables a more synthetic look at the error and simulation times. Clearly, performance may be increased either by improving the accuracy of the method or by achieving lower simulation times. 
\begin{equation}
    \label{eq:performance_eta}
    \eta = \frac{1}{t_{\textrm{sim}} \times \MDErr}   
\end{equation}

We fixed an absolute minimum tolerance $\dQMin = 10^{-9}$ and tested seven values for the relative error tolerance $\dQRel$ between $2.5 \times 10^{-4}$ and $7.5 \times 10^{-3}$. Each scenario was independently simulated 20 times. Sample standard deviations (for simulation times) remained below $12\%$.

\subsubsection{Results and discussion}
\label{sec:md_results}

\begin{figure}[h]
    \centering
    \includegraphics[scale=0.6]{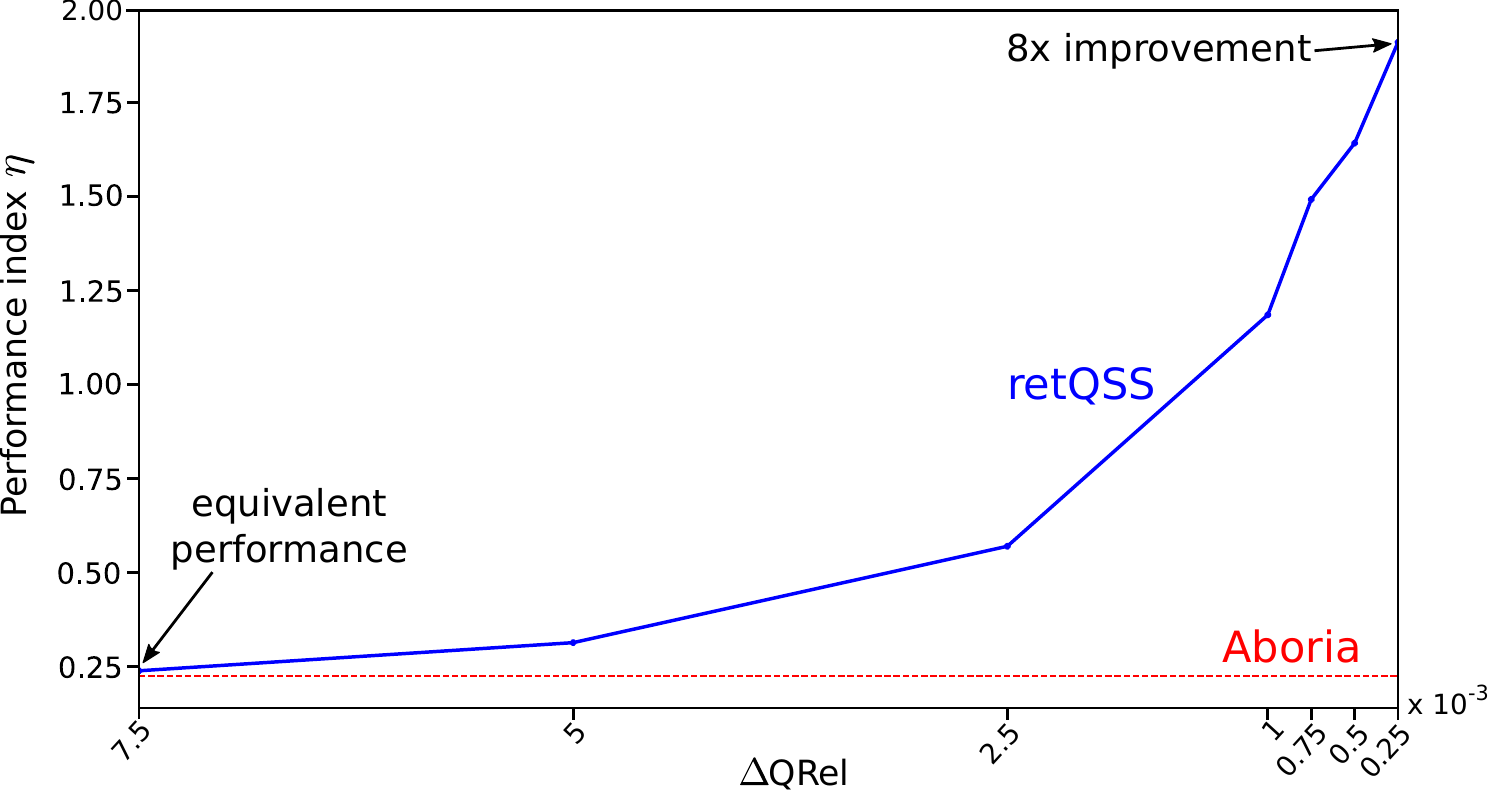}\par 
    \caption{Performance comparison between \retQSS and Aboria}
    \label{fig_md_perf}
\end{figure}

Figure \ref{fig_md_perf} compares the performance index $\eta$ achieved by Aboria and \retQSS. The dashed red line represents the performance index of Aboria (about 0.23, given by a mean simulation time of 1.33 seconds and an error of $3.3 \times 10^{-3}$, i.e., a 0.33\% of the extent of the geometry) whereas the full blue line joins the performance indices achieved by the seven \retQSS simulations configured with increasing accuracy constraints.
 
When $\dQRel = 7.5 \times 10^{-3}$, \retQSS achieves a nearly equivalent performance than Aboria, completing the simulation in 1.38 seconds (about $4\%$ slower) but with an 8\% improvement in the error. From that point onwards, \retQSS  systematically improves its performance as the requested accuracy increases, reaching an improvement of 8x when $\dQRel = 2.5 \times 10^{-4}$. This is due to consistent improvements in the error bounds, a reasonable consequence of more stringent accuracy demands. The decrease in the error follows an expected pattern until $\dQRel = 7.5 \times 10^{-4}$ (i.e., it decreases by factors roughly equivalent to decrease factors in $\dQRel$). From that point onwards, the error decreases more slowly and it seems to stabilize around a value about an order of magnitude less than the error achieved by Aboria.

In a similar fashion, simulation times in \retQSS also grow as the accuracy constraints become more tight. For example, when $\dQRel = 2.5 \times 10^{-4}$, \retQSS completes the simulation in 1.45 seconds (about $9\%$ slower than Aboria). The $\sim$10x improvement in the error bounds clearly overcomes this modest difference, resulting in the reported 8x improvement in the performance index. 

This gentle growth in the simulation time can be explained as follows. First, we confirmed that the number of internal transitions of the QSS2 integration loop follows an expected growth pattern, increasing roughly proportionally to the square root of the increase in the accuracy (as described in Section \ref{sec:qss_properties}). Since the quantum becomes  systematically smaller, these extra transitions are basically due to changes in the state variables. In this case study, these transitions are considerably cheaper than event-driven transitions, which involve particle neighborhood traversals (to compute force fields) and geometrical calculations (to penetrate periodic boundaries). The number of such events is less affected by changes in the requested accuracy (e.g., force field evaluations are only triggered by time events). Thus, the $5\%$ increase in the simulation time (from $\dQRel = 7.5 \times 10^{-3}$ to $\dQRel = 2.5 \times 10^{-4}$) corresponds to a $\sim$5x increase in the number of state changes.
 
\subsubsection{Modeling overview}
\label{sec:md_modeling_overview}

We can highlight several advantages of modeling this case study with \retQSS. As opposed to Aboria, the underlying integration method is decoupled from the modeling logic. The modeler may choose among different options (e.g. QSS2 or QSS3) without worrying about updating the model code. Also, these methods are already tested and deployed in an optimized simulation engine. Another feature of \retQSS is a predictable global accuracy control that is easily configurable (also shown by the performance comparison studied above). In the case of Aboria, the programmer must manually embed accuracy control mechanisms in the model code. Finally, in \retQSS we can easily experiment with geometries of arbitrary shapes, whereas Aboria is limited by design to hypercube geometries. 

Meanwhile, a downside of the \retQSS approach is the need to provide a precise meshed description of the geometry so that particle neighborhoods can properly capture the cut-off value of the force field. In some circumstances, experimenting with different cut-off values may thus demand to recompute the mesh. Also, \retQSS is currently limited to 3D geometries, whereas Aboria supports arbitrary $n$-dimensional hypercubes.

\subsection{Plasma flow}
\label{sec:plasma}

Modeling and simulation of plasma phenomena has become a valuable tool in disciplines such as plasma physics or space physics \cite{Ledvina2008ModelingPhenomena}. In the latter, for example, plasma simulations allow for a better understanding of solar wind interactions with different celestial objects like comets or planets. High-density plasma is typically represented by magnetohydrodynamics (MHD) models, regarded as an extension of fluid dynamics to electrically conducting fluids. On the other hand, low-density plasma tends to behave more like a collection of discrete particles rather than a continuous fluid. Particle-based numerical methods are thus well suited for these problems.

One of the most popular particle-based methods for plasma simulation is Particle-In-Cell (PIC). PIC operates with a background mesh through which particles indirectly interact with each other. In electrostatic setups (e.g. assuming that currents generated by the plasma are low and that the self-induced magnetic field can be ignored), the algorithm follows four basic steps \cite{Markidis2011TheMethod}: 
\begin{enumerate}
    \item The charge of each particle is distributed to the surrounding nodes of the mesh.
    \item Electric potential is calculated by discretizing Poisson's equation and solving the resulting system of linear equations.
    \item Electric forces acting on each particle are computed from the electric field values in the surrounding nodes of the mesh.
    \item Particles are moved using Newton's laws (new particles might be generated after this step).
\end{enumerate}

In what follows we will explore two approaches to implement custom electrostatic PIC algorithms in \retQSS. We tackle a case study that models the flow of uniform plasma in a 2D domain and compare our proposed models against a reference model solved with a PIC implementation in Octave. This programming language was already used in past works to characterize and study PIC algorithms \cite{Konior2017Particle-In-CellAlgorithm}. Bearing in mind the important differences between interpreted and compiled languages, the fundamental purpose of this case study is somewhat different from that of previous experiments. We do not intend to measure and compare the performance of our approach against Octave. Rather, we expect to illustrate the modeling capabilities of our tool offering two PIC variants of very different nature --one of them featuring a novel event-based charge scattering approach that exploits the efficient discrete event capabilities of \retQSS.

\subsubsection{Model description}
\label{sec:plasma_model}

\begin{figure}[H]
    \centering
    \includegraphics[scale=0.13]{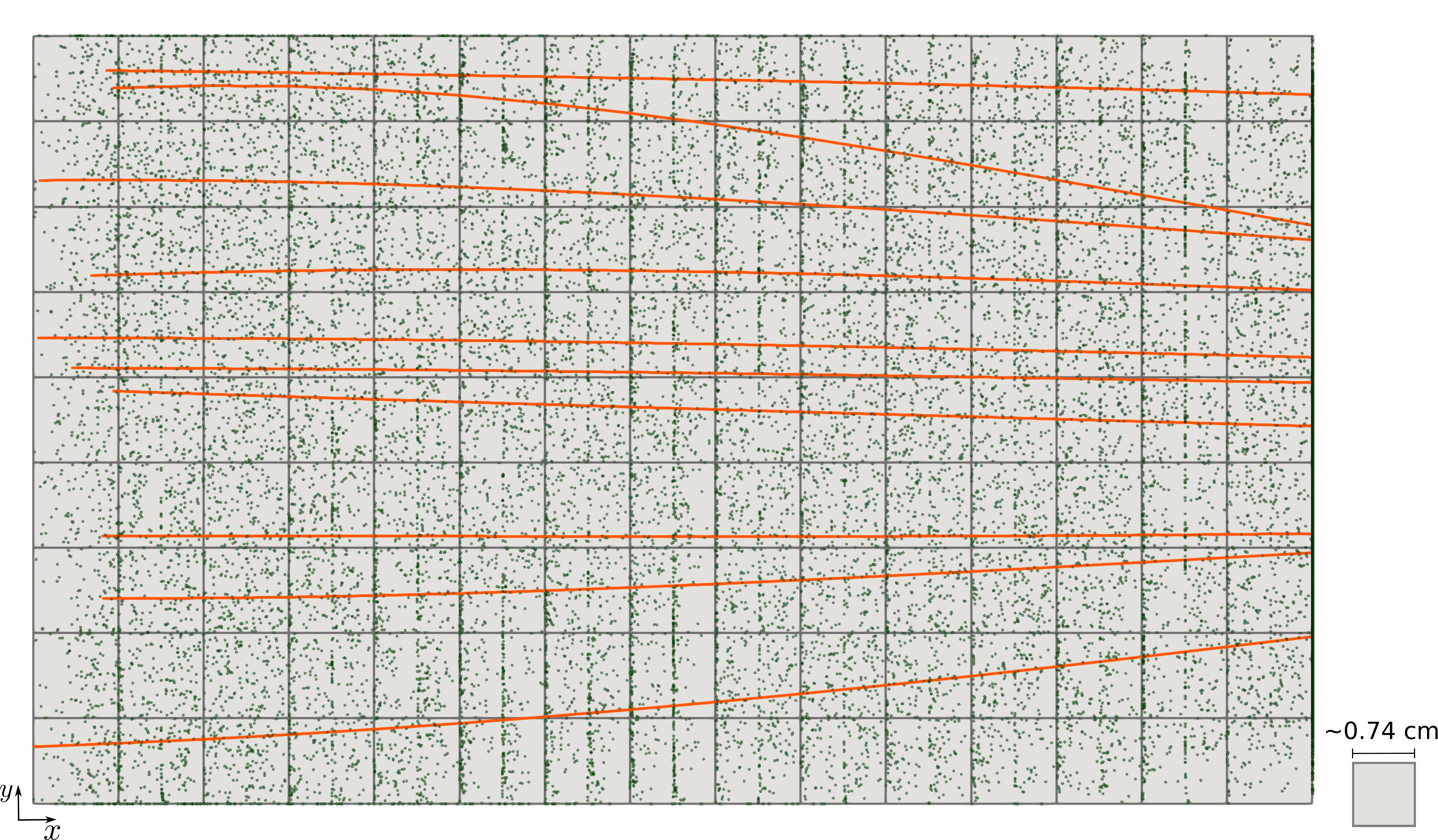}\par 
    \caption{Snapshot of a plasma flow simulation (with highlighted trajectories)}
    \label{fig_plasma}
\end{figure}

We model the flow of uniform plasma in a 2D domain with absorbing boundaries (i.e., particles die when they leave the domain). The background support mesh is given by a grid of $N_x = 16 \times N_y = 10$ nodes, conforming a mesh of square cells where the side length $l$ is given by the Debye length (a characteristic distance over which ions and electrons can be separated in a plasma). At each time step (after $T = 200$ time steps the system reaches a steady state), $n = (N_y-1) \cdot n_p$ macroparticles (computational particles that represent a group of real particles such as electrons or ions) are generated and randomly distributed along the cells of the first column ($n_p$ is a model parameter that controls the average number of macroparticles per cell). The initial velocity of each macroparticle is sampled from the Maxwell-Boltzmann distribution, with a drift factor ($7 \,\, \sfrac{\textrm{km}}{\textrm{s}}$) added in $\hat{x}$. Figure \ref{fig_plasma} shows a snapshot of a simulation with a total of $N = 27000$ macroparticles \blue{(the default value, representing approximately $10^{10}$ real particles).}

The ODE system shown in Equation~\eqref{eq:newton_plasma} dictates the motion of each macroparticle.

\begin{equation}
\systeme{\dot{x} = v_x \quad\quad \dot{v_x} = \sfrac{q_e}{m} \cdot E_x,
          \dot{y} = v_y \quad\quad \dot{v_y} = \sfrac{q_e}{m} \cdot E_y
\label{eq:newton_plasma}}    
\end{equation}\\

Newton's second law determines the acceleration in terms of the elementary charge $q_e = 1.602 \times 10^{-19}$ coulombs, an ion mass given by $m = 32$ AMUs and the electric field $\vec{E} = (E_x, E_y)$, which is computed from $\vec{E} = -\nabla \phi$. Here, $\phi$ is the electric potential, given by Poisson's equation (Equation~\eqref{eq:poisson}). This equation is discretized following a finite difference approach (using central differences).

\begin{equation}
    \nabla^2 \phi = -\frac{\rho}{\epsilon_0} 
\label{eq:poisson}
\end{equation}

Here, $\epsilon_0 = 8.854 \times 10^{-12}$ farads per meter is the permittivity of free space and $\rho$ is the charge density, defined in terms of the densities of ions and electrons. Electron density is given by the Boltzmann relation, whereas ion density is derived from the background mesh after scattering macroparticle charge to its nodes. The charge is distributed to the background mesh using a first-order scattering operation (area weighting). Essentially, a node $v$ surrounding a macroparticle $p$ receives a fraction of the total charge carried by $p$ that is proportional to the distance between $v$ and $p$. 

\subsubsection{Implementation}
\label{sec:plasma_implementation}

We followed two approaches to model this case study in \retQSS. First, we explored a \textsl{standard} strategy where charge distribution is performed on a regular basis, controlled by a \textit{time event} that iterates over every active macroparticle. Then, we tested an alternative \textsl{decoupled} strategy where charge is distributed on demand at two discrete \textit{state events}: when a macroparticle enters into a new cell, and  when a macroparticle is approximately halfway through a cell. Both models share a common general structure and set of features:
\begin{itemize}
    \item Ion charge density and electric potential are stored as vertex properties.
    
    \item The linear system obtained by discretizing Poisson's equation is solved in an external C++ function that provides an ad-hoc implementation of the Gauss-Seidel method. This function is periodically invoked by a time event.
    
    \item The charge scattered by macroparticles is recorded so that it can be removed from previous nodes before scattering it into new ones.
\end{itemize}

The Octave model relies upon an ad-hoc implementation of a leapfrog integrator embedded in the code. In \retQSS, we chose QSS2 as the back-end integration method.

\subsubsection{Validation}
\label{sec:plasma_validation}

We performed 50 independent runs of the reference Octave model in its default configuration (a total of 27000 macroparticles, injecting 135 per time step) and recorded the final charge density across every node in the grid. Using the corresponding initial particle conditions, we computed the ion charge density with our models obtaining the average relative error for each node (we set $\dQRel = 10^{-6}$ and $\dQMin = 10^{-9}$).

For the \textit{Standard} model, the average of the relative errors (across every node) turned out to be approximately $4\%$ ($8\%$ for the 90th percentile). On the other hand, we found an average relative error of $13\%$ for the \textit{Decoupled} model ($55\%$ for the 90th percentile). This is an expected difference derived from the less frequent charge distributions.

\begin{figure}[H]
    \centering
    \includegraphics[scale=0.6]{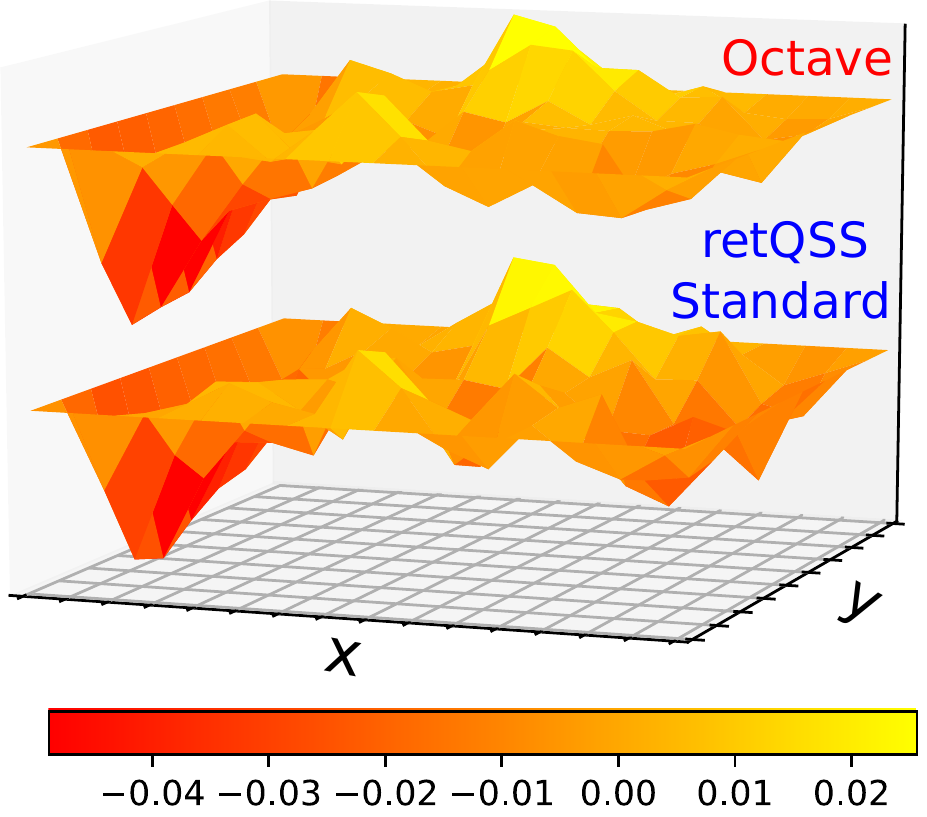}\par 
    \caption{Electric potential for sample simulations in Octave and the Standard model}
    \label{fig_plasma_potential}
\end{figure}

Figure \ref{fig_plasma_potential} shows the final electric potential across the grid computed by the \textit{Standard} model, compared against the one produced by Octave. Although the structure of both surfaces is visually similar, the most remarkable differences occur in the last column, where most of the particles exit the geometry.

\subsubsection{Experiment methodology}
\label{sec:plasma_methodology}

We evaluated our models by testing their performance under different loads of particles. To this end, we swept a range of consecutive values for the $n_p$ parameter ranging from 1 (yielding a total of $N = 1800$ macroparticles) to 15 ($N = 27000$). Each scenario was independently simulated 20 times. Sample standard deviations remained below $5\%$.

\subsubsection{Results and discussion}
\label{sec:plasma_results}

\begin{figure}[h]
    \centering
    \includegraphics[scale=0.6]{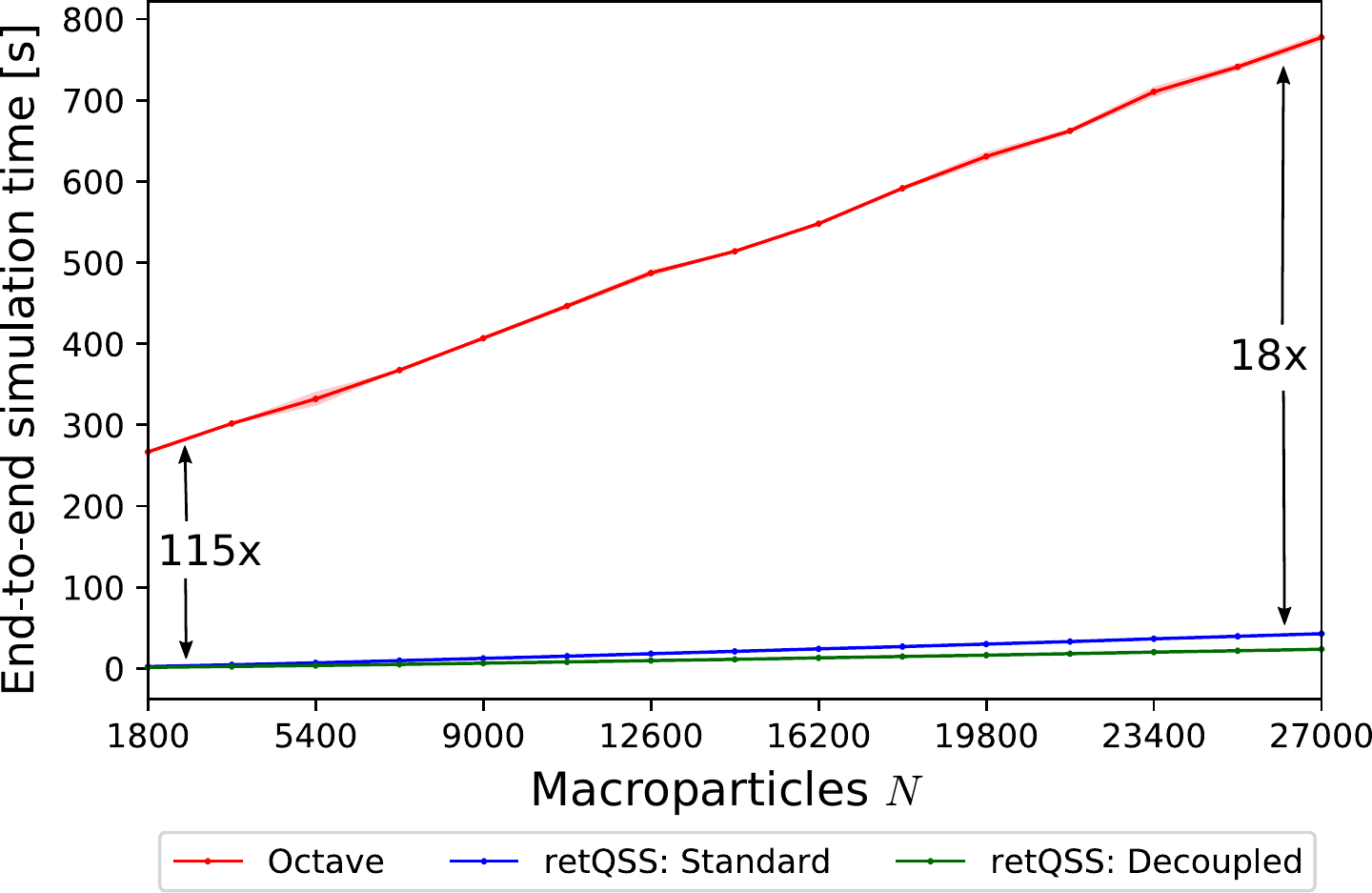}\par 
    \caption{Performance comparison between Octave and \retQSS}
    \label{fig_plasma_perf}
\end{figure}

Figure \ref{fig_plasma_perf} shows that both \retQSS models significantly outperform Octave, featuring speedups of up to 115x (when $N = 1800$). Yet, this difference systematically decreases as the total number of macroparticles increases. The \textit{Standard} model, for example, reduces its speedup against Octave to 18x when $N = 27000$ (the default configuration of the case study).
This can be explained by analyzing the internal event scheduling mechanism used by QSS Solver \blue{(see Algorithm \ref{alg:qss1})} and the extra load induced by the incremental addition of macroparticles to the system. Each new macroparticle adds six continuous state variables and \blue{three to four discrete events}, all with their associated next times of change in the future. These times are managed by the scheduler so that the simulation can advance by processing one state change or discrete event at a time (asynchronous integration steps ordered by their temporal imminence). Thus, the number of such steps (i.e. transitions of the main integration loop) grows linearly with the number of macroparticles $N$.

At each step, the time at which the next transition should happen must be computed. Since we employ the default binary scheduler, the CPU time required by this task is proportional to the logarithm of the total number of states and events. \blue{We also have that, on each step, the number of state derivatives to be recomputed is constant (the underlying mathematical model is sparse). From this analysis we can derive a $\Theta(N \log{N})$ asymptotic bound for the running time of the simulation.} 

\begin{figure}[H]
    \centering
    \includegraphics[scale=0.6]{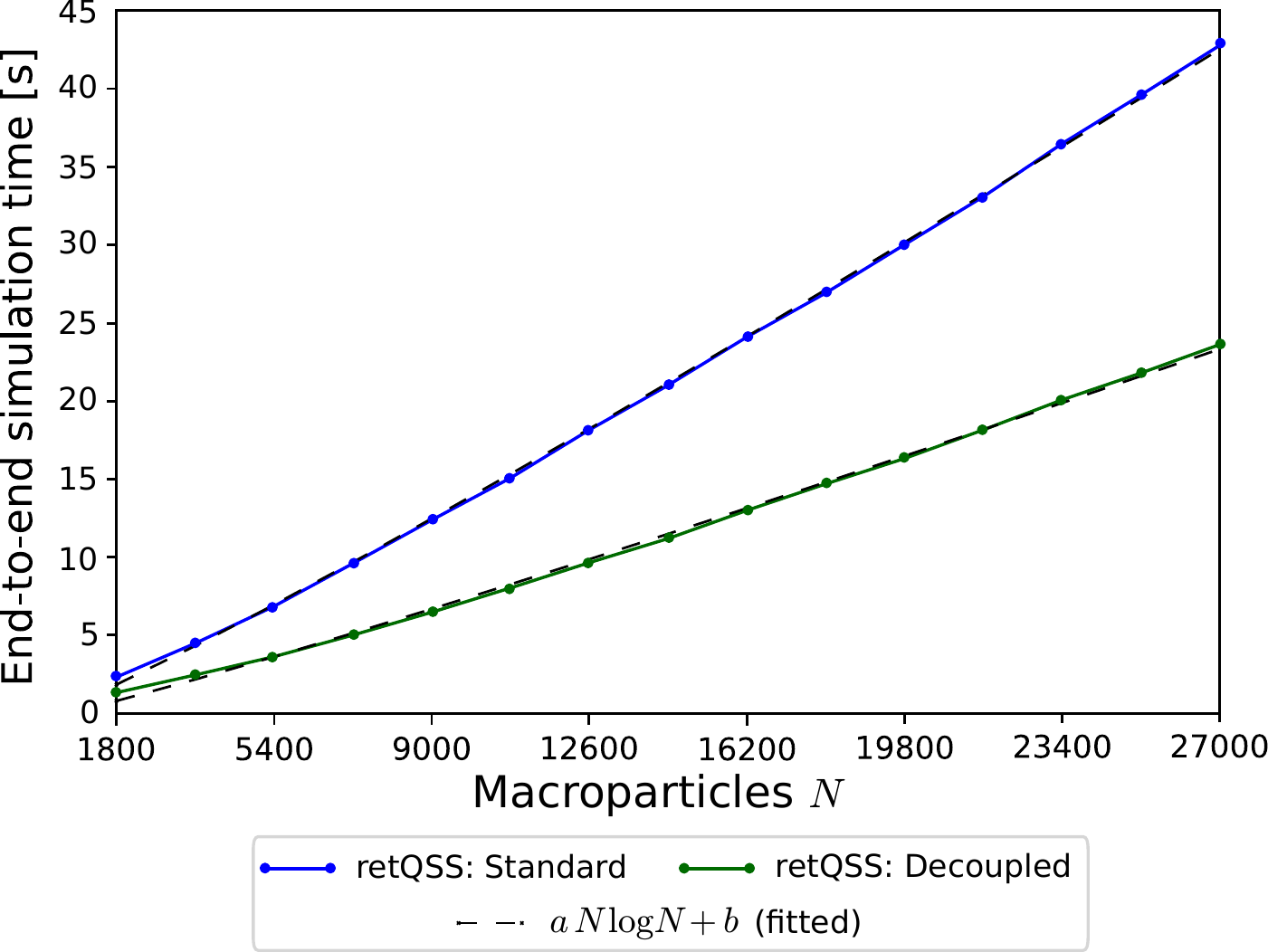}\par 
    \caption{Simulation times for \retQSS strategies compared to fitted functions of the form $a \, N \log N + b$}
    \label{fig:plasma_nlogn}
\end{figure}

\blue{We provide an empirical validation of this assertion in Figure \ref{fig:plasma_nlogn}, where we fit functions of the form $a \, N \log N + b$ to the simulation times yielded by both \retQSS strategies. We also verified that this analysis can be extrapolated to higher values of $N$ (we tested up to $N = 2 \times 10^6$ macroparticles).}

On the other hand, the Octave model simply iterates over the $N$ macroparticles on each time step, yielding an asymptotically linear running time. In consequence, even though both \retQSS models significantly outperform Octave in this setup, this analysis, supported by the empirical evidence, suggests that they scale worse with the number of macroparticles in the system.

Comparing our proposed models, we observe that \textit{Standard} is consistently about $80\%$ slower than \textit{Decoupled}. This is a reasonable outcome, as charge is more frequently distributed to the grid nodes. Yet, if we take into account not only the simulation times but also the average relative errors in the ion charge density, by Equation \eqref{eq:performance_eta} we find that \textit{Standard} outperforms \textit{Decoupled} (achieving a combined performance improvement of $75\%$). The 3x difference in the error is sufficient to compensate for the running time penalties derived from its charge scattering mechanism.
 
\subsubsection{Modeling overview}
\label{sec:plasma_modeling_overview}

The discrete-event nature of \retQSS enables interesting and novel approaches to PIC algorithms, as illustrated by the \textit{Decoupled} model. Sacrificing accuracy in favor of efficiency, this model adjusts the frequency of the charge scattering operation to the dynamics of each macroparticle, a seemingly complex task for traditional approaches such as the Octave model. Still, more advanced models (e.g. featuring splines or other higher-order scattering operations) can be difficult to describe via high-level modeling languages. This might be overcome by developing a custom package offering assorted utility functions targeted to PIC models. These functions can then leverage the versatility of the Model API in their implementations.

\section{Related work and discussion}
\label{sec:related_work}

This Section aims at contextualizing \retQSS to understand how it is positioned within a reference framework of particle simulation methods and tools. To this end, we shall review relevant contributions to this broad topic in a selection of specific disciplines: computational fluid dynamics (CFD), high-energy physics (HEP) and 3D rendering.

\subsection{Computational fluid dynamics}
\label{sec:related_work_cfd}

CFD models often involve the simultaneous flow of fluids and particles. These coupled models employ discrete particle simulation methods to obtain the phase motion of the particles in the system, usually applying Newton's laws of motion to each particle. One such method is the Discrete Element Method (DEM) \cite{OSullivan2011ParticulatePerspective}. DEM is an example of mesh-free, particle-based numerical methods. Particle methods employ a finite set of discrete particles to represent the state of a system \cite{Li2007MeshfreeMethods}. Particles are then tracked in a Lagrangian frame according to their internal interactions and external forces, and thus evolve the system across time. Some particle methods are best suited for systems with discrete particles (e.g., rocks, grains or atoms) whereas others are used to discretize continuous domains such as fluids. DEM and Molecular Dynamics (MD) \cite{Rapaport2004TheSimulation} are two classic examples of the former. Smoothed Particle Hydrodynamics (SPH) \cite{Gui-rong2003SmoothedMethod} is the most widely used method among the latter. A central aspect common to all of them is that they approximate forces acting on a particle using information of neighboring particles lying inside a \textsl{support domain} \cite{Gui-rong2003SmoothedMethod}. As we covered in Section \ref{sec:md}, the related concept of particle neighborhood enables implementations of MD with \retQSS.

Other class of particle methods operate using a background Eulerian mesh. Although considered a mesh-free particle method, Particle-In-Cell (PIC) is one such example \cite{Birdsall2004PlasmaSimulation}. Originally limited to fluids, PIC later evolved into other methods that expanded its capabilities (e.g. the Material Point Method, MPM \cite{Idelsohn2018ParticleDynamics}). In these methods, particle properties (e.g. electric charge) are interpolated in the surrounding mesh nodes on each simulation step. Later, fields are interpolated back to particles after computing their value over the nodes. As we studied in Section \ref{sec:plasma}, the use of meshed geometries in \retQSS is a key feature that paves the way for custom implementations of these sorts of particle methods. We covered a case study involving the flow of plasma in a simple two-dimensional geometry and developed two different PIC-based approaches to model this problem in \retQSS.

A related class of hybrid methods used in CFD are those that combine Lagrangian and Eulerian grids so as to leverage the advantages of both approaches. Two notorious examples are the Coupled Eulerian-Lagrangian (CEL) method \cite{Gui-rong2003SmoothedMethod} and the Arbitrary Lagrangian-Eulerian (ALE) method \cite{Margolin1997IntroductionSpeeds}. CEL usually applies Eulerian and Lagrangian methods in separate regions of the problem domain (e.g., solids and fluids may be discretized in Lagrangian and Eulerian frames, respectively). Both regions continuously interact with each other through a coupling module in which computational information is exchanged. An important difference with \retQSS (and in general with the aforementioned particle-based methods) is that these are grid-based approaches that do not make explicit use of computational particles.

One of the most popular software toolkits for CFD is OpenFOAM \cite{Jasak2007OpenFOAM:Simulations}. It is an open-source, C++-based toolbox for general-purpose CFD simulations. OpenFOAM supports discrete particle simulations via custom implementations of DEM. MercuryDPM \cite{Weinhart2020FastMercuryDPM} is another example of discrete particle simulation toolkits. One of its salient features, not available in \retQSS, is the support of curved walls that might arise in realistic geometries. On the other hand, Nauticle \cite{Havasi-Toth2020Nauticle:Tool} offers a custom high-level language (the Symbolic Form Language, SFL) that facilitates the formulation of user-defined numerical models. Although similar in spirit to \retQSS, this mechanism is not based on a widely accepted, standardized modeling language such as Modelica. Yet another example is Aboria \cite{Robinson2017Particle-basedAboria}, an efficient software library aimed at providing a common framework for the implementation of particle-based methods (see Section \ref{sec:md} for detailed comparisons between \retQSS and Aboria).
 
\subsection{High-energy physics}
\label{sec:related_work_hep} 
 
The HEP community relies heavily on particle tracking simulations. These serve multiple purposes, e.g. they drive the design and optimization of particle detectors for best physics performance \cite{Elvira2017DetectorSimulation}. MARS15 \cite{Mokhov2007MARS} is a HEP particle simulator consisting of a set of Monte Carlo algorithms for the simulation of hadronic and electromagnetic cascades in 3D geometries. FLUKA \cite{Bohlen2014FLUKA} is a general-purpose simulation package for particle transport and interactions with matter. Yet, modern HEP experiments mostly rely upon Geant4 \cite{Allison2016Geant4} for carrying out particle simulations. Geant4 is a software toolkit for the simulation of the passage of particles through matter. Its main goal is to track the trajectories of subatomic particles affected by physics processes within complex detector geometries typically composed by 3D volumes of different shapes and materials. In Section \ref{sec:hep} we compared \retQSS and Geant4 in the context of a simple HEP setup (a charged particle moving in a mesh of cuboids under the action of a uniform magnetic field). We developed variants of this setup that were successfully used in the past as a benchmarking asset for QSS-based simulation strategies ported to Geant4 \cite{Santi2019APhysics,Santi2020HEP}.

Geant4 employs a custom, XML-based geometry file format (GDML, Geometry Description Markup Language), through which it supports geometries with complex features such as nested volumes. As a consequence, the underlying geometry management and tracking algorithms are considerably more complicated than those featured by \retQSS. Thus, advanced HEP setups featuring complex geometries (such as realistic particle detectors) may prove difficult to model in \retQSS. One way of overcoming this limitation is to adopt a  co-simulation strategy to connect Geant4 with QSS Solver, as we introduced in \cite{Santi2019APhysics}. This approach opens the door to leveraging the benefits of \retQSS in other application domains.

\subsection{3D rendering}
\label{sec:related_work_3d}

In the field of 3D rendering and game physics, the Particle Systems method \cite{Reeves1983ParticleObjects} is generally used to model fuzzy objects such as fire or water, difficult to reproduce with standard rendering techniques. This method tracks a particle cloud in a 3D space, advancing particle positions (and possibly other features) by means of a physical simulation which might consider external forces like gravity or friction. In a very similar fashion to \retQSS simulations, the method usually detects collisions between the particles and other objects in the scene, enabling different kinds of interactions with 3D obstacles. Although particle systems solutions are sometimes developed ad-hoc, companies like Nvidia \cite{Nvidia} provide Particle Systems APIs that can be used in game engines.

\subsection{Event-driven particle dynamics and particle tracking}
\label{sec:related_work_edpd}

Another particle simulation technique very closely linked to \retQSS and with application in a wide range of fields is Event-Driven Particle Dynamics (EDPD) \cite{Bannerman2014StableDynamics}. EDPD maps the dynamics of a particle system to a sequence of events given by instantaneous pairwise interactions between the particles, thus progressing irregularly in time. EDPD is in consequence a discrete-event system, much like the QSS numerical methods. Nevertheless, EDPD is typically used in hard-sphere models with colliding particles. These event-driven particle-particle collisions are not yet supported by \retQSS.

The algorithmic approach followed by EDPD is related to several particle tracking algorithms that have been applied in different fields such as CFD, granular flow simulations and ray tracing \cite{Strobl2020RobustGeometries,Macpherson2009ParticleDynamics}. In this context, particle tracking refers to the repeated localization of particles within a mesh by following their trajectories. Indeed, this is one of the central tasks of \retQSS. Although the boundary crossing detection algorithm implemented by \retQSS  (covered in Section \ref{sec:crossing_algorithm}) has a strong resemblance to the aforementioned tracking strategies, these are generally based upon linear approximations of particle trajectories. On the other hand, \retQSS relies on higher order polynomial approximations of the trajectories supplied by QSS Solver, ultimately enabling more accurate estimations of the intersection points.

\section{Conclusions and future work}
\label{sec:conclusions}

This work introduced \retQSS, a novel methodology for modeling and simulation of particle systems in meshed three-dimensional geometries. Aimed at providing a generic, rigorous and efficient methodological framework for particle simulation, retQSS adopts and extends the QSS Solver toolkit as its core simulation engine. This way, \retQSS profits not only from its elegant mechanisms to describe models (via \uModelica, a high-level modeling language that is a subset of the Modelica language) but also from its optimized, state-of-the-art implementations of QSS numerical methods. Due to their discrete-event nature, these methods enable efficient approaches to particle tracking algorithms, a key component in particle simulations across different application domains. 

We demonstrated the modeling capabilities of \retQSS by addressing four selected case studies. First, we explored two different approaches to model bird flocking as suggested by the very popular \textsl{boids} model. The concept of particle neighborhoods in \retQSS enabled straightforward implementations of the flocking rules that dictate the steering behavior of birds according to their local flockmates. We also discussed how the model can be easily extended to take into account external forces (e.g. wind) or the presence of obstacles, leveraging other features of \retQSS such as volume properties.

Then, we continued with a high-energy physics (HEP) setup based on a model used in our previous contributions as a benchmarking asset for QSS-based simulation strategies for HEP experiments. A performance comparison against Geant4, the reference simulation toolkit in the field, revealed substantial speedups (between 6x and 8x) in detecting and handling volume boundary crossings, a central aspect in HEP simulations. This proved the efficiency and appropriateness of our discrete-event based particle tracking algorithms. Yet, extending this analysis to advanced HEP setups (such as realistic models of particle detectors) may prove challenging, as these models typically involve more complex geometries (e.g. featuring nested volumes).

Finally, we showed how particle-based numerical methods can be effectively expressed in \retQSS. We developed a custom implementation of Molecular Dynamics in the context of a system of molecules interacting through an exponentially decaying potential, and compared our approach against Aboria, a software library devoted to particle-based methods. We exploited the straightforward accuracy control mechanisms in \retQSS to improve the error bounds in about one order of magnitude with limited performance penalties (simulation times resulted 9\% higher than those achieved by Aboria). In addition, we showed two different strategies to implement Particle-In-Cell algorithms to model the flow of plasma in a 2D domain. One of them explored a novel discrete-event based approach to the problem of scattering particle charge to the computational grid. Although our approaches greatly outperformed a reference implementation in Octave (reaching speedups of up to 115x), they scaled worse with the number of particles in the system. 

We believe that the results of this work position \retQSS as a powerful new alternative to other related particle simulation toolkits. From a modeling perspective, by seizing the expressive power of a high-level modeling language, \retQSS features succinct and elegant model descriptions, something that may prove particularly useful for disciplines typically accustomed to using programming languages for modeling purposes. 

We are currently working on enabling parallel computation mechanisms in \retQSS, leveraging the shared-memory approach followed by QSS Solver. We are also interested in coupling \retQSS with other simulation engines such as OpenModelica. This would allow for tackling a wider range of problems by using the full capabilities of Modelica to describe particle models.


\blue{To conclude, interesting future steps include extending \retQSS with convenient primitives to facilitate domain-specific modeling. One such domain is the modeling of \emph{continuous contacts}, which is especially relevant for instance in particle-wall interactions in DEM models. Another area of interest is the relationship between particle-based and agent-based models for studying systems involving social interactions such as epidemiological processes. In this branch of modeling, agents are represented by particles following kinetic laws while describing situations of encounter and contagion \cite{kuzdeuov2021,pulvirenti2020}. We started exploring this line of work in \cite{lanzarotti2021} by extending \retQSS to build an agent-based model that combines the kinematic 2D motion of agents, indirect interaction between them and with their surrounding space, and centralised control to apply contact tracing over the entire population.}

\section*{Acknowledgments}

This work was supported by the National Agency for Science and Technology (ANPCYT, grant PICT-2015-3509) and the University of Buenos Aires (UBACYT PhD Fellowship Program).

\appendix

\section{Hardware and software platform}
\label{sec:platform}

All simulations (single-threaded) were run on the computer cluster TUPAC \cite{TUPAC2}, where each CPU node has 4 x AMD Opteron 6276 (hexadeca-core) processors. The operating system in use is Red Hat Enterprise Linux ComputeNode release 6.7\\ (\texttt{2.6.32-573.el6.x86\_64} kernel).

Regarding the third-party software employed throughout the experimentation,
\begin{itemize}
    \item The Geant4 version in use was 10.05, released December, 2018.
    
    \item As for Aboria, we used version 0.5, the latest official release as of the writing of this article.
    
    \item Finally, we used Octave version 4.2.1, released February, 2017.
\end{itemize}

With the exception of Octave, every piece of software (including \retQSS and QSS Solver) was compiled from the source code with \texttt{gcc} 5.4.0 and enabling the \texttt{O2} optimization flag. The QSS Solver version adopted for the development of \retQSS is a fork from version 3.0.

\bibliography{misc}

\end{document}